\documentclass[11pt]{article}

\usepackage{geometry}                
\usepackage{graphicx,amssymb,bbm,mathrsfs,cite,epstopdf}
\def\met{\mbox{${\hbox{$E$\kern-0.6em\lower-.1ex\hbox{/}}}_T$}}
\begin{document}
\vspace*{0.5cm}

\begin{center}	
\thispagestyle{empty}

{\LARGE\bf  Interpreting LHC SUSY searches in the\\[3mm] phenomenological MSSM}\\[12mm]

{\large S. Sekmen$^{\,a}$, S.~Kraml$^{\,b}$, J.~Lykken$^{\,c}$, F.~Moortgat$^{\,d}$, S.~Padhi$^{\,e}$,\\[2mm] L.~Pape$^{\,d}$, M.~Pierini$^{\,f}$, H.\,B.~Prosper$^{\,a}$, M.~Spiropulu$^{\,f,g}$}\\[6mm]

{\it
$^{a}$~Department of Physics, Florida State University,\\ Tallahassee, Florida 32306, USA\\[2mm]
$^{b}$~Laboratoire de Physique Subatomique et de Cosmologie, UJF Grenoble 1, 
CNRS/IN2P3, INPG, 53 Avenue des Martyrs, F-38026 Grenoble, France\\[2mm]
$^{c}$~Fermi National Accelerator Laboratory, P.O. Box 500, Batavia, IL 60510\\[2mm]
$^{d}$\,Institute for Particle Physics, ETH Zurich, CH-8093 Zurich, Switzerland\\[2mm]
$^{e}$\,University of California, San Diego, CA 92093, USA\\[2mm]
$^{f}$\,CERN, CH-1211 Geneva 23, Switzerland\\[2mm]
$^{g}$\,Department of Physics, Caltech, Pasadena, California 91125, USA\\[8mm]
}

\vspace*{6mm}

\begin{abstract}
We interpret within the phenomenological MSSM (pMSSM) the results of SUSY 
searches published by the CMS collaboration based on the first $\sim$1~fb$^{-1}$ of data 
taken during the 2011 LHC run at 7~TeV. 
The pMSSM is a 19-dimensional parametrization of the MSSM that captures 
most of its phenomenological features. It encompasses, and goes beyond, 
a broad range of more constrained SUSY models. 
Performing a global Bayesian analysis, we obtain  
posterior probability densities of parameters, masses and derived observables.
In contrast to constraints derived for particular SUSY breaking schemes,  
such as the CMSSM, our results provide more generic conclusions on how the current 
data constrain the MSSM.
\end{abstract}

\end{center}
\clearpage

\section{Introduction}
%
With the successful operation of the Large Hadron Collider (LHC) and its detectors
in 2010--11, and with excellent prospects for the future, the LHC is ready to shed light 
on the most pressing open issues in particle physics: 
the mechanism of electroweak~(EW) 
symmetry breaking and the nature of the new physics 
beyond the Standard Model that stabilizes the EW scale 
(see, for example, Ref.~\cite{Giudice:2007qj}).

A wealth of theories that extend the Standard Model~(SM) have been put forth 
during the past decades. 
Among these, supersymmetry (SUSY) is arguably the best motivated and certainly the 
most thoroughly studied (see, for example,  Refs.~\cite{Martin:1997ns,Chung:2003fi} for recent reviews). 
Indeed, searches for SUSY rank among the primary experimental objectives of the LHC. 
So far, however, no signal of new physics has been observed at $\sqrt{s}=7$~TeV~\cite{lp11}; 
consequently, the SUSY mass scale has been pushed into the TeV region.  

It is important to note, however, that in the interpretation of their 
experimental results, both the ATLAS and CMS collaborations typically  
use a very special theoretical model, the so-called Constrained Minimal 
Supersymmetric Standard Model (CMSSM), which is characterized by just four-and-a-half parameters~\cite{Chamseddine:1982jx,cmssm}: 
a universal scalar mass $m_0$, 
gaugino mass $m_{1/2}$ and trilinear coupling $A_0$  
defined at the GUT scale $M_{\rm GUT}\sim 10^{16}$~GeV, plus 
$\tan\beta$ and $\rm{sign(\mu)}$. 
The simplifying assumption of universality at the GUT scale 
makes the model very predictive and a convenient showcase for SUSY phenomenology. 
Indeed, it is interesting to present limits within the CMSSM because it provides 
(to some degree) an easy way to show performances, compare limits or reaches, 
etc..\ On the other hand, the interpretation of experimental results in the 
$(m_0,m_{1/2})$ plane risks imposing unwarranted constraints on SUSY, as many 
mass patterns and signatures that are possible \emph{a priori} are not covered 
in the CMSSM. 
The same of course holds true for Simplified Models~\cite{Alves:2011wf}. 

In this Letter, we therefore present a more general approach, using a 
19-dimensional parametrization of the MSSM 
called the \emph{phenomenological MSSM} (pMSSM)~\cite{Djouadi:1998di}. 
Using results from three independent CMS analyses---the 
$\alpha_T$~hadronic ~\cite{Collaboration:2011zy}, 
the same-sign dilepton~\cite{CMS:SS} and 
the opposite-sign dilepton~\cite{CMS:OS} analyses---we derive constraints 
on the SUSY particles with as few simplifying assumptions as possible. 

The pMSSM parameter space has been thoroughly scanned and studied 
previously in Refs.~\cite{Berger:2008cq,AbdusSalam:2009qd,Conley:2010du,Conley:2011nn}. 
It is important to note that~\cite{Berger:2008cq}  
{\it ``the pMSSM leads to a much broader set of predictions for the properties 
of the SUSY partners as well as for a number of experimental observables than 
those found in any of the conventional SUSY breaking scenarios such as mSUGRA [CMSSM]. 
This set of models can easily lead to atypical expectations for SUSY signals 
at the LHC.''}

The purpose of this Letter is thus to initiate a systematic study that begins with 
an assessment of 
what current LHC data tell us, and do not tell us, about the pMSSM. 
We choose to conduct our study 
using the Bayesian approach~\cite{Bayes:1, Bayes:2} because of its  conceptual coherence and the
direct (intuitive) manner in which
probabilistic statements are interpreted, namely,  as the degree of belief, or
plausibility, of a given statement. 
A detailed Bayesian study of the pMSSM was performed in Ref.~\cite{AbdusSalam:2009qd}; however this was before LHC data were available.\footnote{In 
Ref.~\cite{Farina:2011bh}, the CMSSM gluino--squark mass limits based on 
1~fb$^{-1}$ of LHC data were applied to a dark matter global fit in 
a 9-parameter realization of the MSSM.} 

We introduce the pMSSM and its parametrization in Section~\ref{sec:model}, 
and outline our analysis in Section~\ref{sec:method}. 
Our results are presented in Section~\ref{sec:results}. 
Section~\ref{sec:conclusions} contains our conclusions.

\section{Parametrization}\label{sec:model}

The pMSSM, a 19-dimensional realization~\cite{Djouadi:1998di} 
of the R-parity conserving MSSM with parameters defined at the SUSY scale,  
$M_{\rm SUSY}=\sqrt{m_{\tilde t_1}m_{\tilde t_2}}$, employs only a few 
plausible assumptions motivated by experiment:
there are no new CP phases, the sfermion mass matrices and trilinear couplings are flavor-diagonal, 
the first two generations of sfermions are degenerate and their trilinear couplings are negligible. 
In addition, we assume that the lightest supersymmetric 
particle (LSP) is the lightest neutralino, $\tilde\chi^0_1$. 
We thus arrive at a proxy for the MSSM characterized by 19 real, weak-scale, 
SUSY Lagrangian parameters:
\begin{itemize}
   \item the gaugino mass parameters $M_1$, $M_2$, and $M_3$; 
   \item the ratio of the Higgs VEVs $\tan\beta=v_2/v_1$;
   \item the higgsino mass parameter $\mu$ and 
            the pseudo-scalar Higgs mass $m_A$;
    \item 10 sfermion mass parameters $m_{\tilde{F}}$, where 
         $\tilde{F} = \tilde{Q}_1, \tilde{U}_1, \tilde{D}_1, 
                      \tilde{L}_1, \tilde{E}_1, 
                      \tilde{Q}_3, \tilde{U}_3, \tilde{D}_3, 
                      \tilde{L}_3, \tilde{E}_3$\\ 
(imposing $m_{\tilde{Q}_1}\equiv m_{\tilde{Q}_2}$, 
           $m_{\tilde{L}_1}\equiv m_{\tilde{L}_2}$, etc.), and          
   \item 3 trilinear couplings $A_t$, $A_b$ and $A_\tau$\,,               
\end{itemize}
in addition to the SM parameters.  

For each pMSSM point, we use  
{\tt SoftSUSY3.1.6}~\cite{Allanach:2001kg} to compute the SUSY spectrum,
{\tt SuperIsov3.0}~\cite{Mahmoudi:2008tp} to compute the low-energy constraints,
{\tt micrOMEGAs2.4}~\cite{Belanger:2001fz}  for the SUSY mass limits, and
{\tt HiggsBounds2.0.0}~\cite{Bechtle:2011sb}  for the limit on the $h^0$ mass\footnote{In 
evaluating the Higgs mass limit, we apply a Gauss-distributed theoretical uncertainty 
with $\sigma=1.5$~GeV to the $m_h$ computed by with {\tt SoftSUSY}, 
cf.\ row 8 in Table~\ref{tab:PLMs}.}. 
Moreover, we use 
{\tt SUSYHIT (SDECAY1.3b, HDECAY3.4)}~\cite{Djouadi:2006bz} to produce SUSY and Higgs decay tables, and
{\tt micrOMEGAs2.4}~\cite{Belanger:2001fz} to compute the LSP relic density and direct dection cross sections.
The various codes are interfaced using the SUSY Les Houches Accord~\cite{Skands:2003cj}.

\section{Analysis}\label{sec:method}

\begin{table}[t]
\begin{center}
\caption{The preLHC experimental results that are the basis of  our pMSSM parameter scan
using Markov Chain Monte Carlo (MCMC) sampling.   
We re-weight a~posteriori with the new limit $BR(B_s \rightarrow \mu \mu) \le 1.08 \times 10^{-8}$ 
at 95\% CL \cite{newbsmm}. However, this has hardly any  effect.}
\begin{tabular}{|c|c|c|c|c|}
\hline
$i$ 	& Observable 	& Experimental result 	& Likelihood function \\
 	& $\mu_i$		& $D_i$				&  $L(D_i|\mu_i)$ \\
\hline\hline
1 & $BR(b \rightarrow s\gamma)$  \cite{Asner:2010qj,Misiak:2006zs}  
   & $(3.55 \pm 0.34)\times 10^{-4}$  
   & Gaussian \\
\hline
2 & $BR(B_s \rightarrow \mu \mu)$ \cite{Nakamura:2010zzi}
   & $\le 4.7 \times 10^{-8}$    
   & $1/\big(1+{\mathrm{exp}}{(\frac{\mu_2-D_2}{0.01D_2})}\big)$ \\
\hline
3 & $R(B_u \rightarrow \tau \nu)$  \cite{Nakamura:2010zzi}
   & $1.66\pm 0.54$  
   & Gaussian \\
\hline
4 & $\Delta a_\mu$ \cite{Davier:2010nc}
   & $(28.7\pm8.0)\times 10^{-10}$ $[e^+e^-]$ & Weighted Gaussian average  \\
   & & $(19.5\pm8.3)\times 10^{-10}$ [taus] &   \\
\hline 
5 & $m_t$ \cite{tev:2009ec}
   & $173.3\pm1.1$~GeV 
   & Gaussian \\
\hline
6 & $m_b(m_b)$  \cite{Nakamura:2010zzi}
   & $4.19^{+0.18}_{-0.06}$~GeV 
   & Two-sided Gaussian  \\
\hline
7 & $\alpha_s(M_Z)$ \cite{Amsler:2008zzb}
   & $0.1176\pm 0.002$ 
   & Gaussian  \\
\hline\hline
8 & $m_h$ & LEP\&Tevatron & $L_8=1$ if allowed. $L_8=10^{-9}$ if   \\
   &             & (HiggsBounds \cite{Bechtle:2011sb}) & $m_h^{'}$ sampled from $Gauss(m_h, 1.5)$  \\
   &             &                          & is excluded.  \\
\hline
9 & sparticle & LEP & $L_9=1$ if allowed  \\
   & masses  & (micrOMEGAs \cite{Belanger:2001fz}) & $L_9=10^{-9}$ if excluded  \\
\hline
\end{tabular}\label{tab:PLMs} 
\end{center}
\end{table}

As noted in the Introduction, 
the purpose of this Letter is to assess  
what current data tell us, and do not tell us, about the pMSSM. 
It is convenient to 
partition these data into preLHC and LHC experimental results, which we
list in Tables~\ref{tab:PLMs} and~\ref{tab:dbgcount}, respectively. 
We use the former to construct a 
prior $\pi(\theta)$ on the pMSSM parameter space,
which, when combined with a likelihood function, $L({\rm LHC}|\theta)$, pertaining to the 
LHC results, yields the posterior density 
$p(\theta|{\rm LHC}) \sim L({\rm LHC}|\theta) \, \pi(\theta)$ over the pMSSM parameter space. Here,
$\theta$ denotes the 19 pMSSM parameters $M_1, \cdots, A_{\tau}$. 
We also consider the SM parameters  
$m_t$, $m_b(m_b)$ and $\alpha_s(M_Z)$, which are treated as nuisance parameters
(see Table~\ref{tab:PLMs}).
This partitioning allows us to assess the impact of
the current LHC results on the pMSSM parameter space 
while being consistent with other existing constraints.

\begin{table}[t]
\caption{LHC measurements used in the current study. 
The $\alpha_T$ variable is effective
in suppressing background from light-quark QCD. SS $2\ell$, and OS $2\ell$ denote same-sign
and opposite-sign dileptons, respectively. The $\alpha_T$~\cite{Collaboration:2011zy}, 
SS~\cite{CMS:SS}, and OS~\cite{CMS:OS} results were published
by the CMS Collaboration.}
\begin{center}
\begin{tabular}{|c|l|c|c|}
\hline
$j$&~Analysis and search region & Observed    & Data-driven SM \\
         &~ (values in GeV)              & event count & BG estimate      \\
&             & $(N_j)$        & $(B_j \pm \delta B_j)$ \\
\hline\hline
1&~$\alpha_T$ hadronic, $275 \le H_T < 325$~ & $782$ & $787.4^{+31.5}_{-22.3}$ \\
2&~$\alpha_T$ hadronic, $325 \le H_T < 375$~ & $321$ & $310.4^{+8.4}_{-12.4}$ \\
3&~$\alpha_T$ hadronic, $375 \le H_T < 475$~ & $196$ & $202.1^{+8.6}_{-9.4}$ \\
4&~$\alpha_T$ hadronic, $475 \le H_T < 575$~ & $62$ & $60.4^{+4.2}_{-3.0}$ \\
5&~$\alpha_T$ hadronic, $575 \le H_T < 675$~ & $21$ & $20.3^{+1.8}_{-1.1}$ \\
6&~$\alpha_T$ hadronic, $675 \le H_T < 775$~ & $6$ & $7.7^{+0.8}_{-0.5}$ \\
7&~$\alpha_T$ hadronic, $775 \le H_T < 875$~ & $3$ & $3.2^{+0.4}_{-0.2}$ \\
8&~$\alpha_T$ hadronic, $875 \le H_T$ & 1 & $2.8^{+0.4}_{-0.2}$ \\
\hline
9&~SS $2\ell$, $H_T>400$, $\met >120$ & $1$ & $2.3 \pm 1.2$ \\
\hline
10&~OS $2\ell$, $H_T>300$, $\met >275$ & $8$ & $4.2 \pm 1.3$ \\
\hline
\end{tabular}\label{tab:dbgcount}  
\end{center}
\end{table}

The prior $\pi(\theta)$ is constructed as follows. We construct the joint likelihood function of the seven independent preLHC measurements $D \equiv  D_1, \cdots, D_7$, of the associated  
observables $\mu \equiv \mu_1,\cdots, \mu_7$,  listed in Table~\ref{tab:PLMs}. 
From Bayes theorem, with  a flat prior $\pi(\mu) = constant$~\footnote{Note that for a Gaussian density, the reference prior is flat (see, for example, Ref.~\cite{Demortier:2010} and references therein).
We will comment on prior dependence in the results section.} for each of the seven observables, we  obtain the posterior density $p(\mu|D) = L(D|\mu) \, \pi(\mu) / p(D)$ from
which we create a random sample of $1.5\times10^7$ pMSSM parameter points using
a standard MCMC technique.
During the sampling, we impose  the constraints on the mass, $m_h$, of the light neutral Higgs boson (given in row 8 of Table~\ref{tab:PLMs}) and the SUSY mass limits (row 9). 
Moreover,  as we cannot scan over an infinite volume, we restrict the sampling to the sub-space
$|M_i|,\, |\mu|, m_A,\, m_{\tilde{F}} \le 3$~TeV, $|A_{t,b,\tau}|\le 7$~TeV, 
and $2\le\tan\beta\le 60$.\footnote{Evidently, for quantities that are not well bounded by the data within the chosen sub-space, the probabilities we calculate will be somewhat sensitive to the choice of sub-space.}
The MCMC sampling from $p(\mu|D)$ together with the predictions $\mu_i = f_i(\theta)$ induce a distribution over $\theta$ that we take as our prior over $\theta$.  
By construction,  the resulting set of pMSSM points are automatically consistent with the preLHC 
experimental constraints listed in Table~\ref{tab:PLMs}.


From the $1.5\times10^7$ Markov-chain points we draw a subset of $5\times10^5$ points, for each of 
which we generate 10K events using {\tt PYTHIA6}~\cite{Sjostrand:2006za}. 
(We checked that both the original chains and their subsets had converged.)  We simulate the response of the CMS detector using the publicly available general purpose detector simulation package {\tt Delphes}~\cite{Ovyn:2009tx}.  
Note, that for studies of this scope, a fast, accurate, detector simulation is essential.  
Regarding LHC results, we use the following three published CMS SUSY analyses:
\begin{itemize}
\item the $\alpha_T$ hadronic analysis~\cite{Collaboration:2011zy}, based on 1.1~fb$^{-1}$,
	$\geq 2$ jets and $\alpha_T > 0.55$, where $\alpha_T$ is used to
	suppress light-flavor QCD, and 8 disjoint bins in $H_T$, the scalar
	sum of jet transverse momenta;
\item the same-sign (SS) di-lepton analysis~\cite{CMS:SS}, based on 0.98~fb$^{-1}$, 
        with 8 overlapping analysis regions of which we use one, $H_T > 400$~GeV,  
        missing transverse energy $\met> 120$~GeV, and
\item the opposite-sign (OS) di-lepton analysis~\cite{CMS:OS}, based on 0.98~fb$^{-1}$,
	with 2 overlapping analysis regions of which we use one, $H_T > 300$~GeV and $\met > 275$~GeV.
\end{itemize}
We take the observed event counts and background estimates directly from the official results of these analyses.  
For each of the ten results listed in Table~\ref{tab:dbgcount},  we assume a Poisson likelihood,
\begin{equation}
  \textrm{Poisson}(N_j | s_j + b_j),
\end{equation}
with observed count $N_j$ and expected count $s_j + b_j$, where $s_j$ and $b_j$ are the 
expected signal and background counts, respectively, for the $j^{\textrm{th}}$ experimental
result.\footnote{We use lower-case letters for parameters and upper-case letters for measured quantities.} 
Each pMSSM point yields predictions for the values $s_j, j = 1,\cdots, 10$. We 
model the (evidence-based) prior for the background parameters $b_i$ with a gamma density,
\begin{equation}
   \textrm{gamma}(K_j b_j|Q_j + 1) = e^{-K_j b_j} (K_j b_j)^{Q_j} / \Gamma(Q_j + 1)\,.
\end{equation}
Here $Q_j \equiv (B_j/\delta B_j)^2$ and $K_j \equiv B_j / \delta B_j^2$, with $B_j \pm \delta B_j$ 
the background estimate in which $\delta B_j$ is taken to be half the width of the confidence intervals 
listed in Table~\ref{tab:dbgcount}. 
For each pMSSM point, and for each result listed in Table~\ref{tab:dbgcount}, we compute
the (marginal) likelihood $p(N_j|s_j)$ by integrating over the expected background $b_j$. Since,
by construction, the results are disjoint, the overall LHC likelihood $L({\rm LHC}|\theta)$ is simply the
product
\begin{equation}
   L({\rm LHC}|\theta) = \prod_{j=1}^{10} p(N_j|s_j(\theta)).
\end{equation}   
The posterior density $p(\theta|{\rm LHC})\propto L({\rm LHC}|\theta) \, \pi(\theta)$ is approximated by
weighting each pMSSM point by $L({\rm LHC}|\theta)$. Finally, we normalize the posterior density over
the pMSSM sub-space.

\section{Results}\label{sec:results}

\begin{figure}[p]
   \centering
   \includegraphics[width=3.5cm]{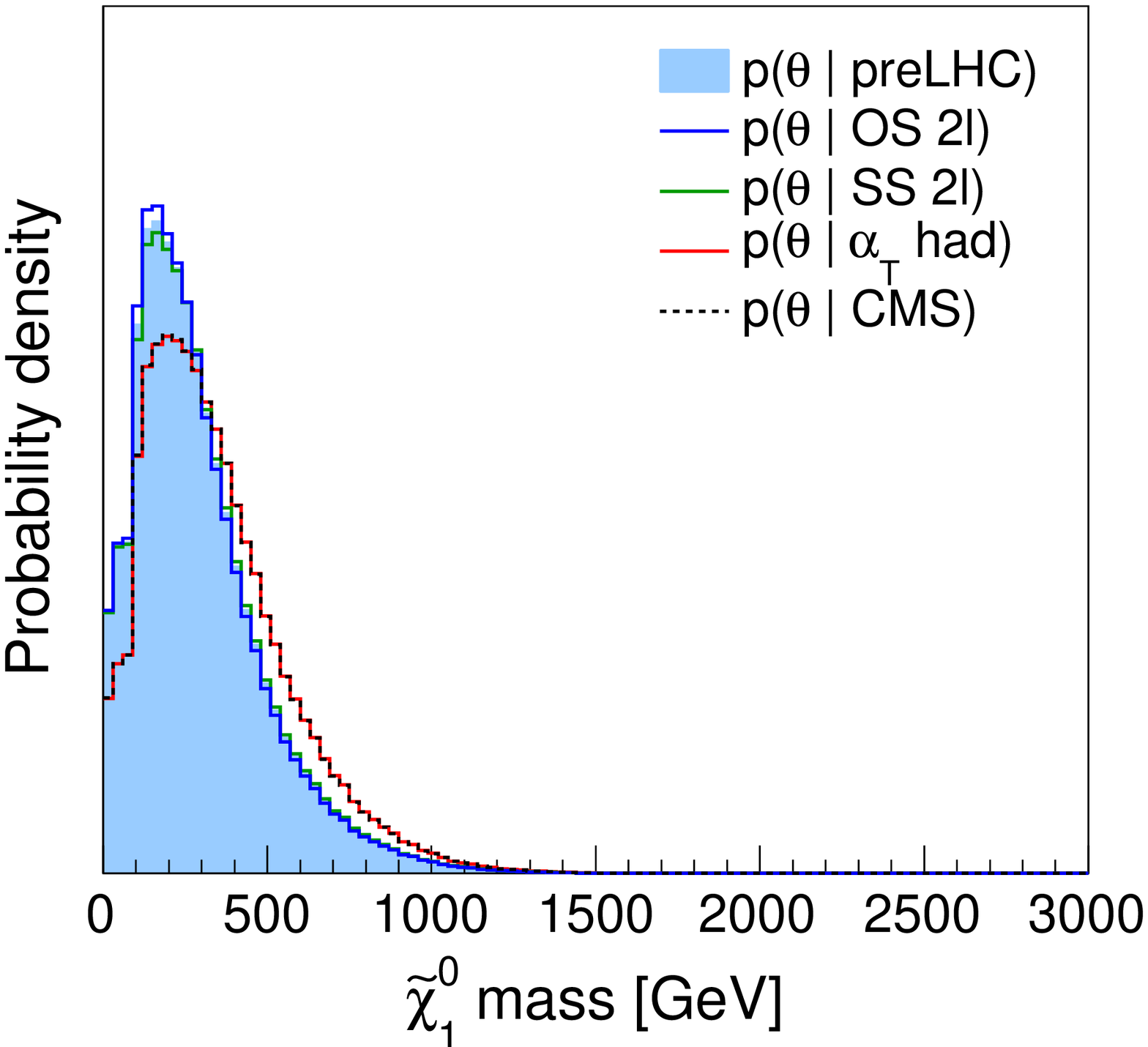} 
   \includegraphics[width=3.5cm]{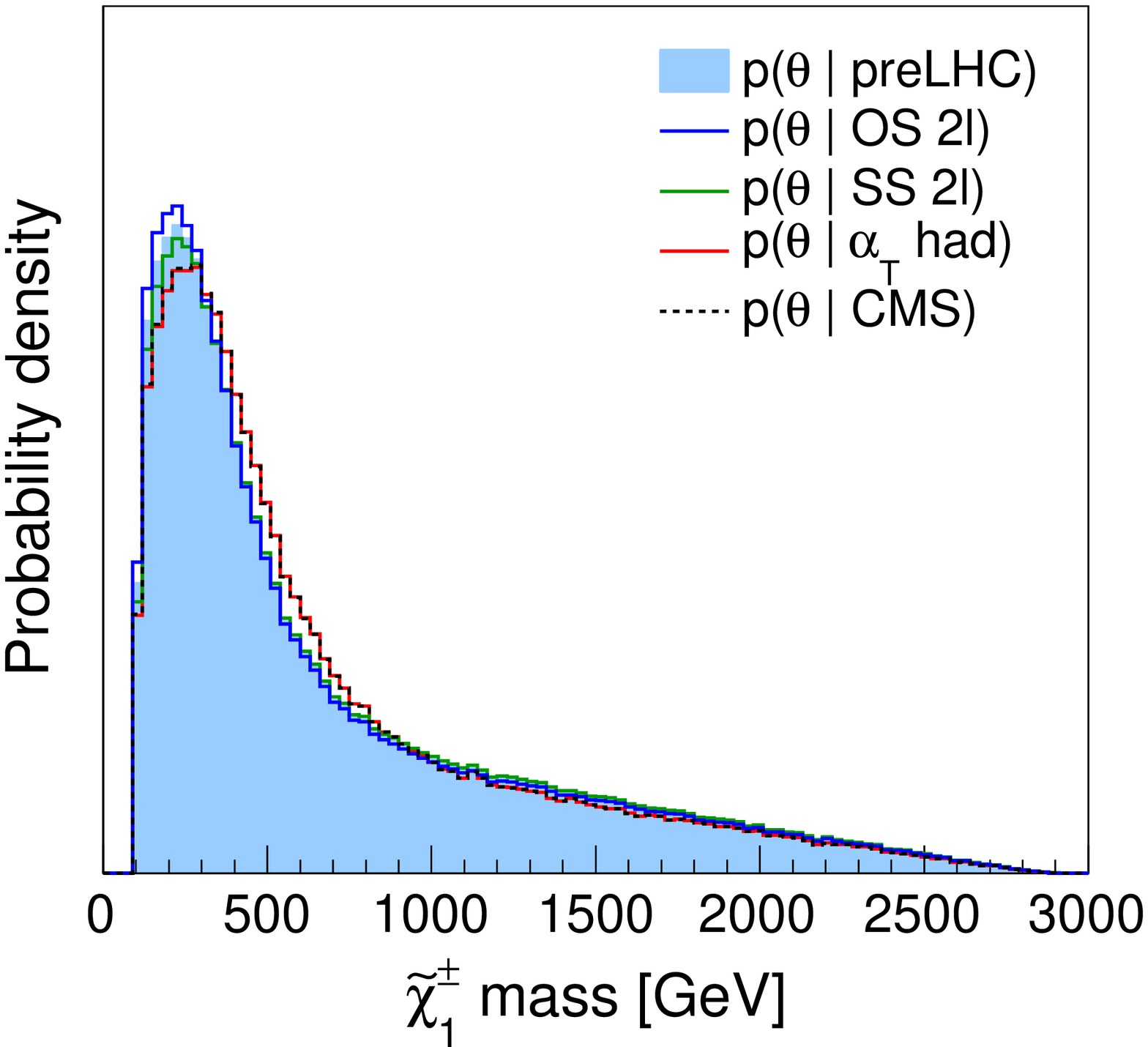} 
   \includegraphics[width=3.5cm]{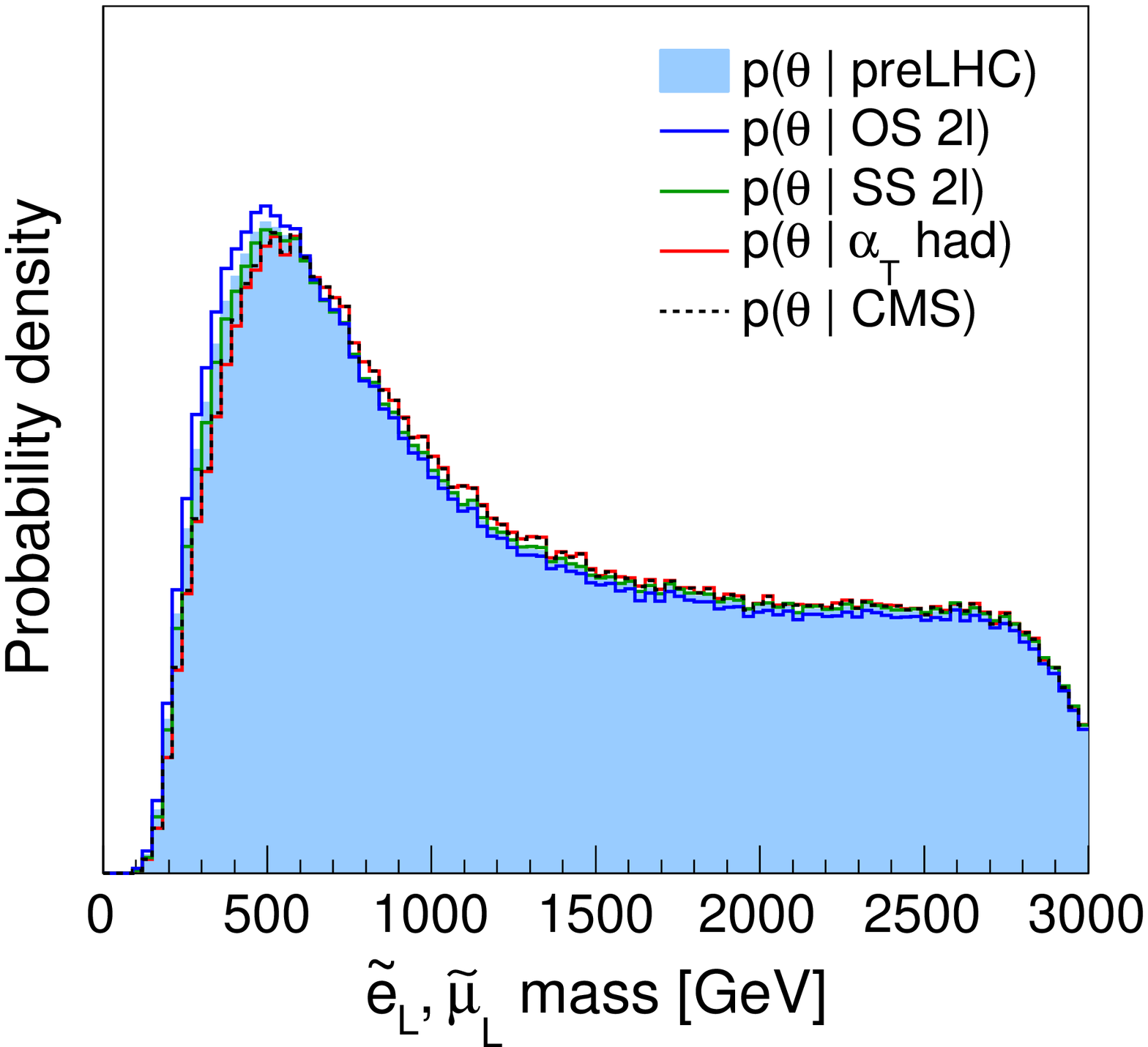} 
   \includegraphics[width=3.5cm]{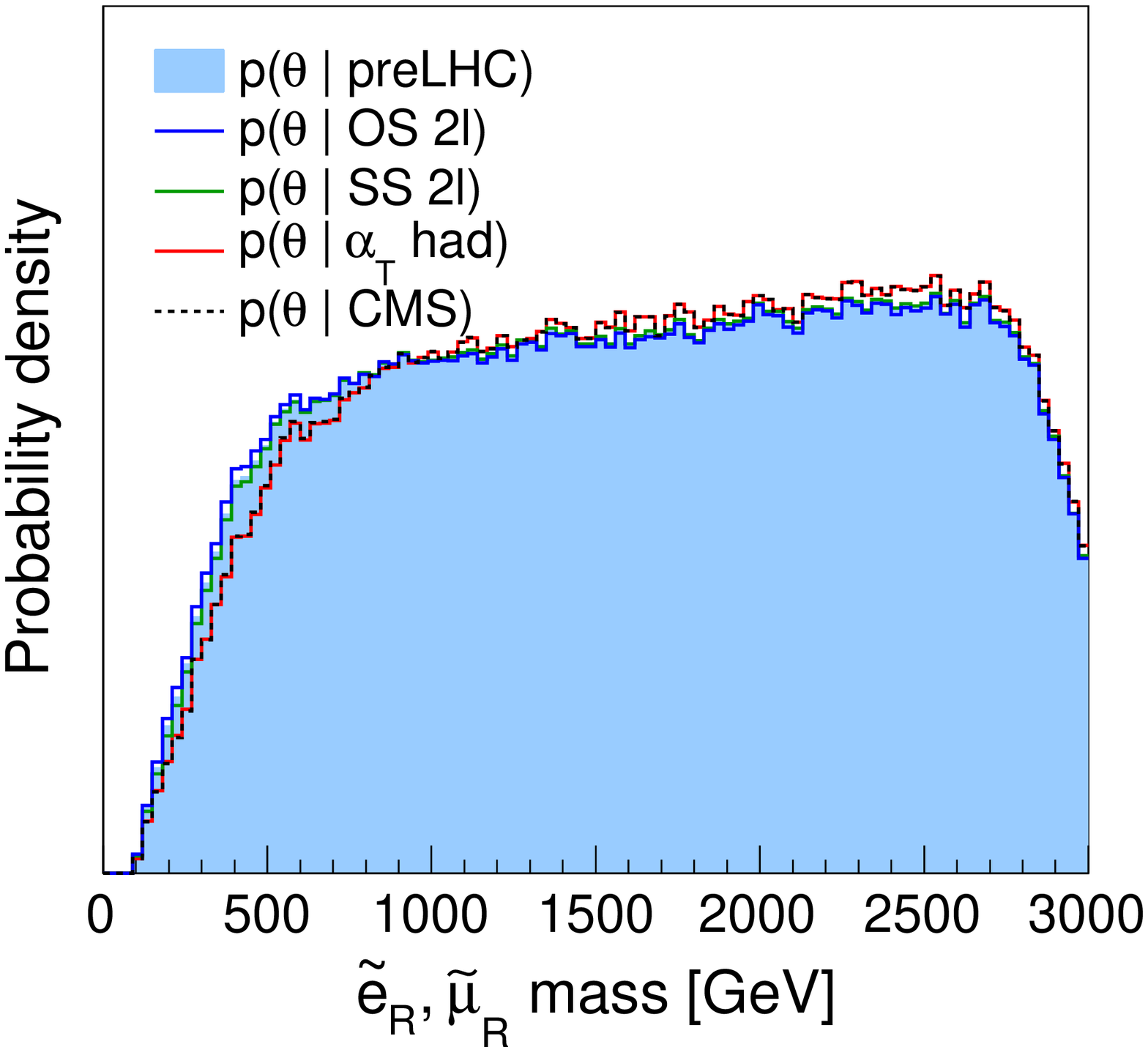} 
   \includegraphics[width=3.5cm]{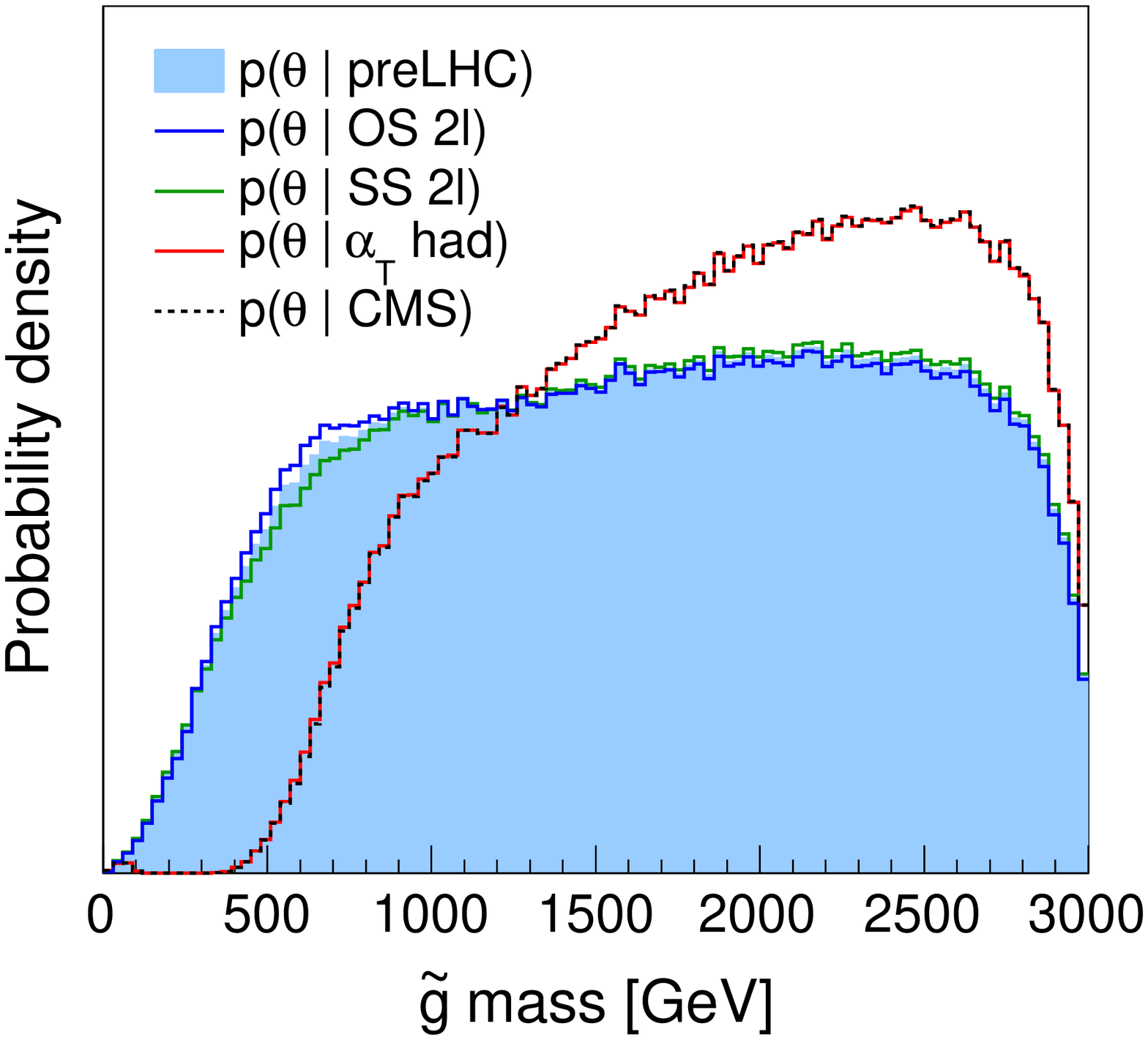} 
   \includegraphics[width=3.5cm]{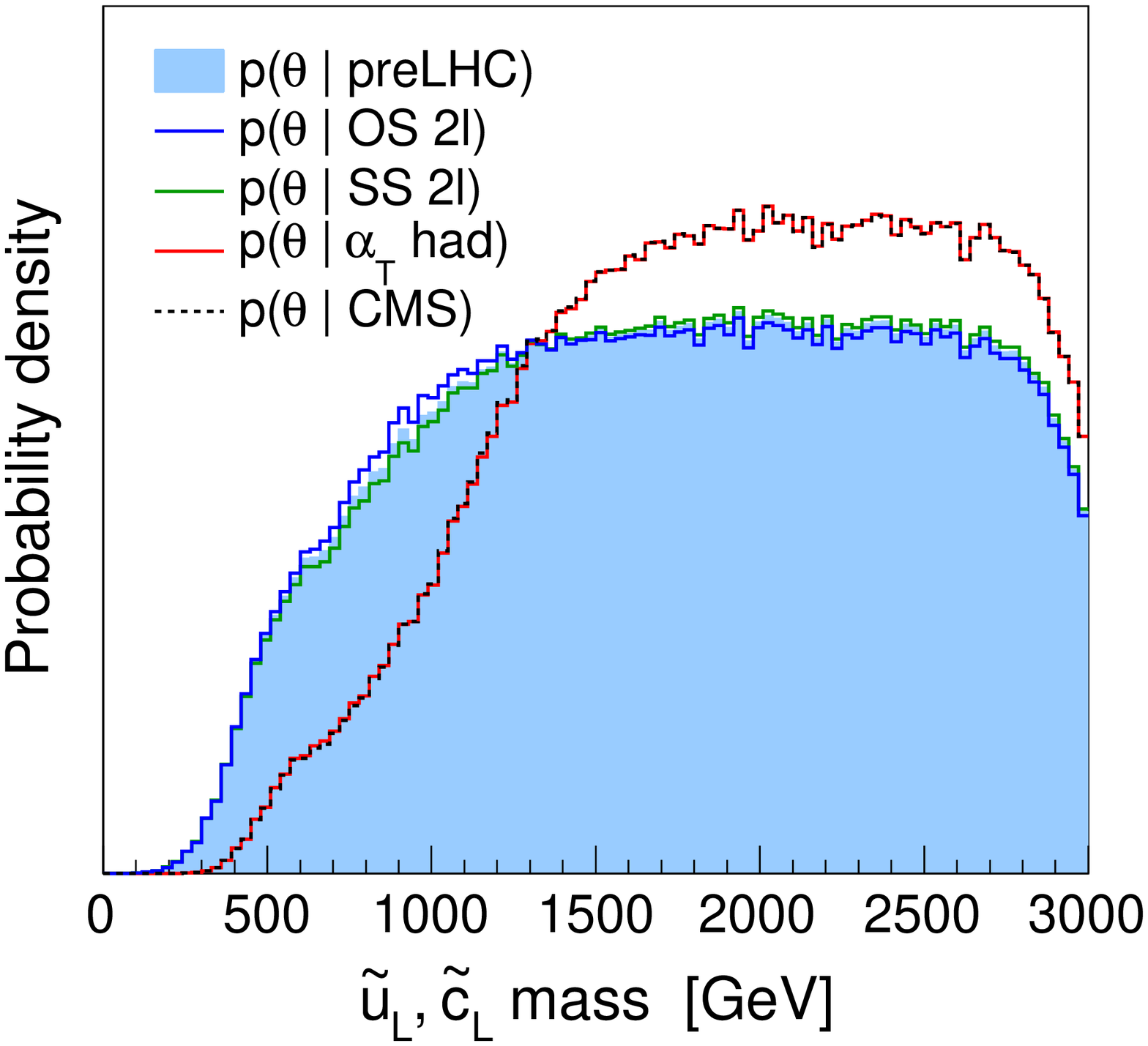} 
   \includegraphics[width=3.5cm]{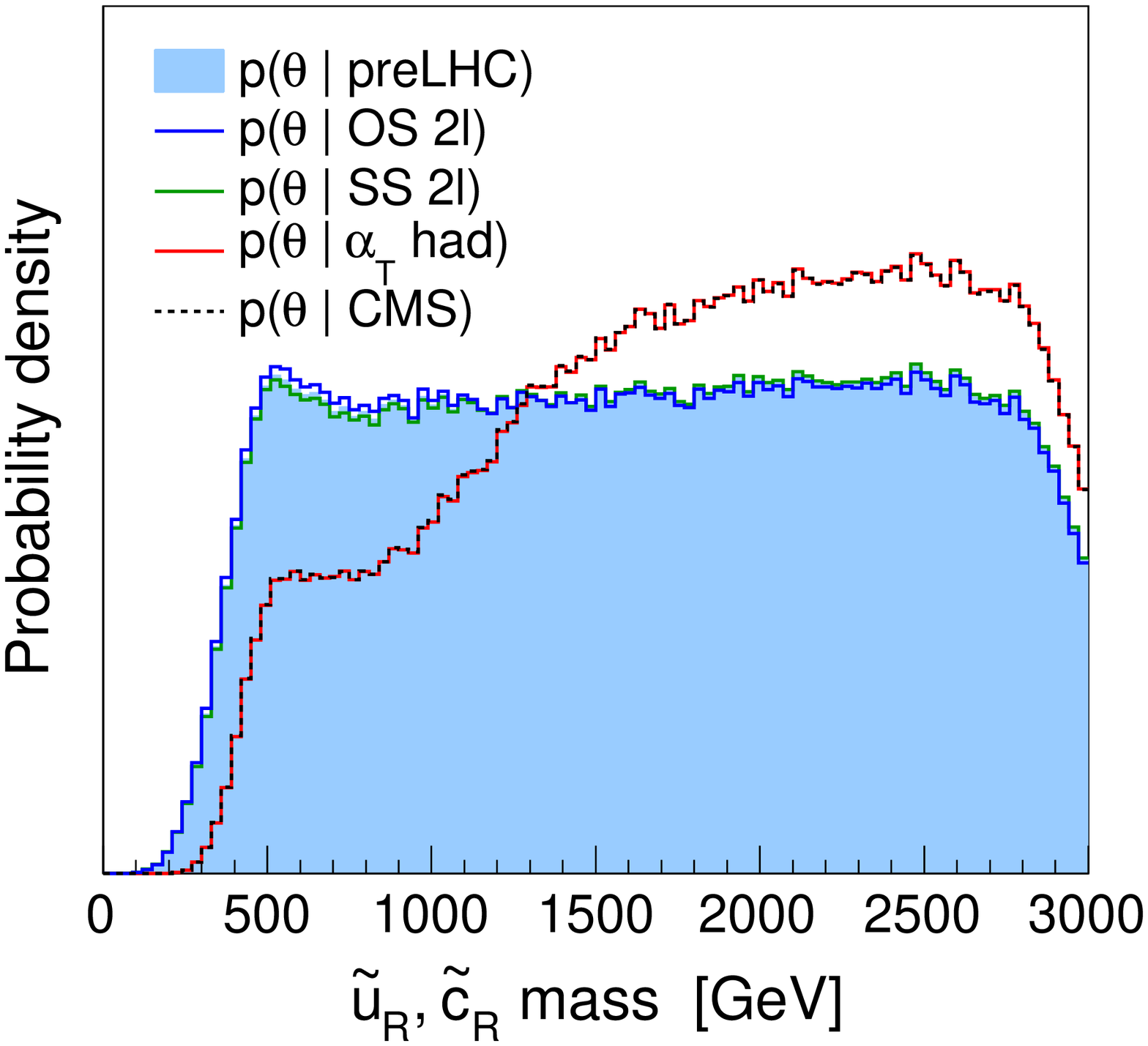} 
   \includegraphics[width=3.5cm]{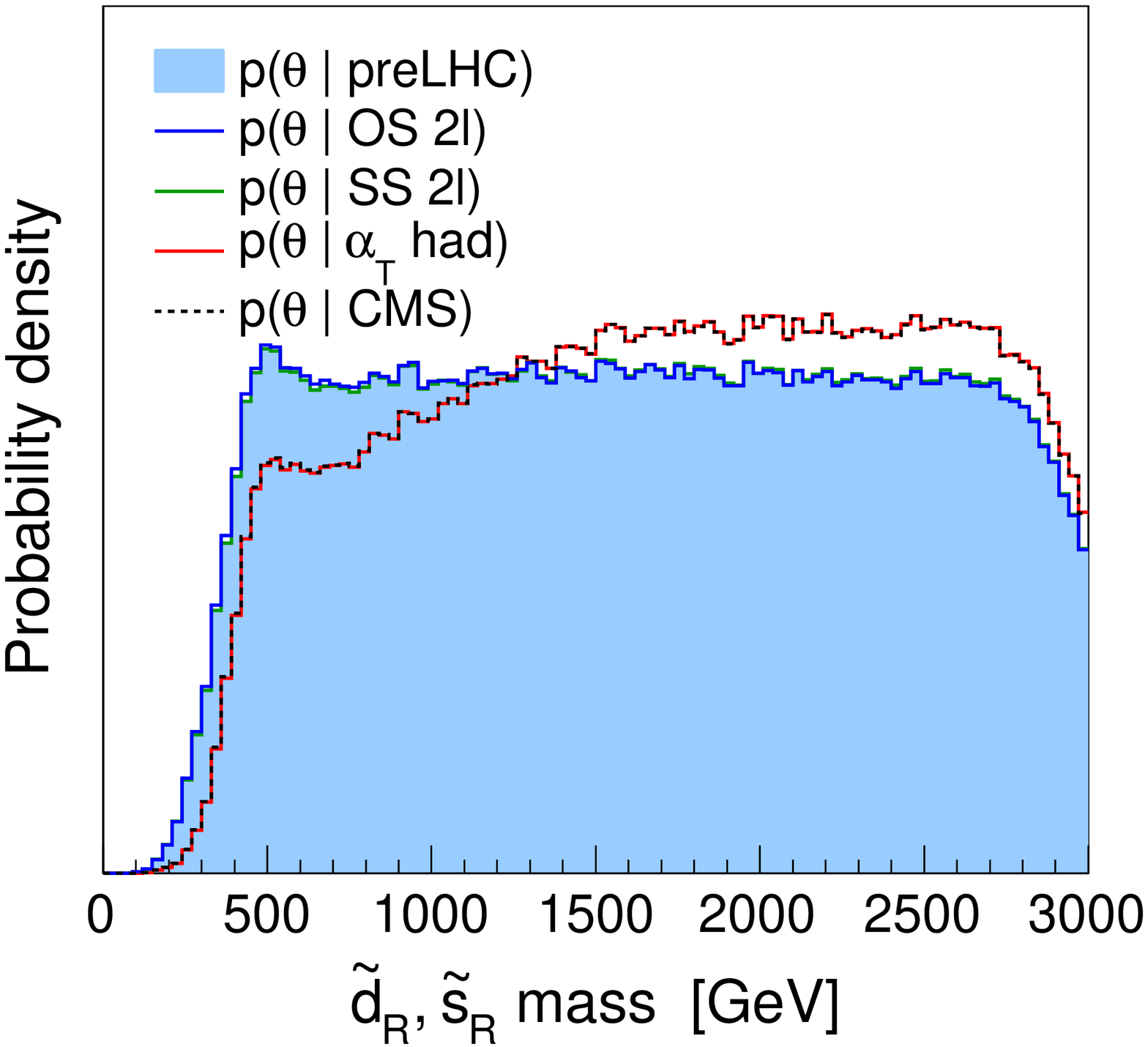} 
   \includegraphics[width=3.5cm]{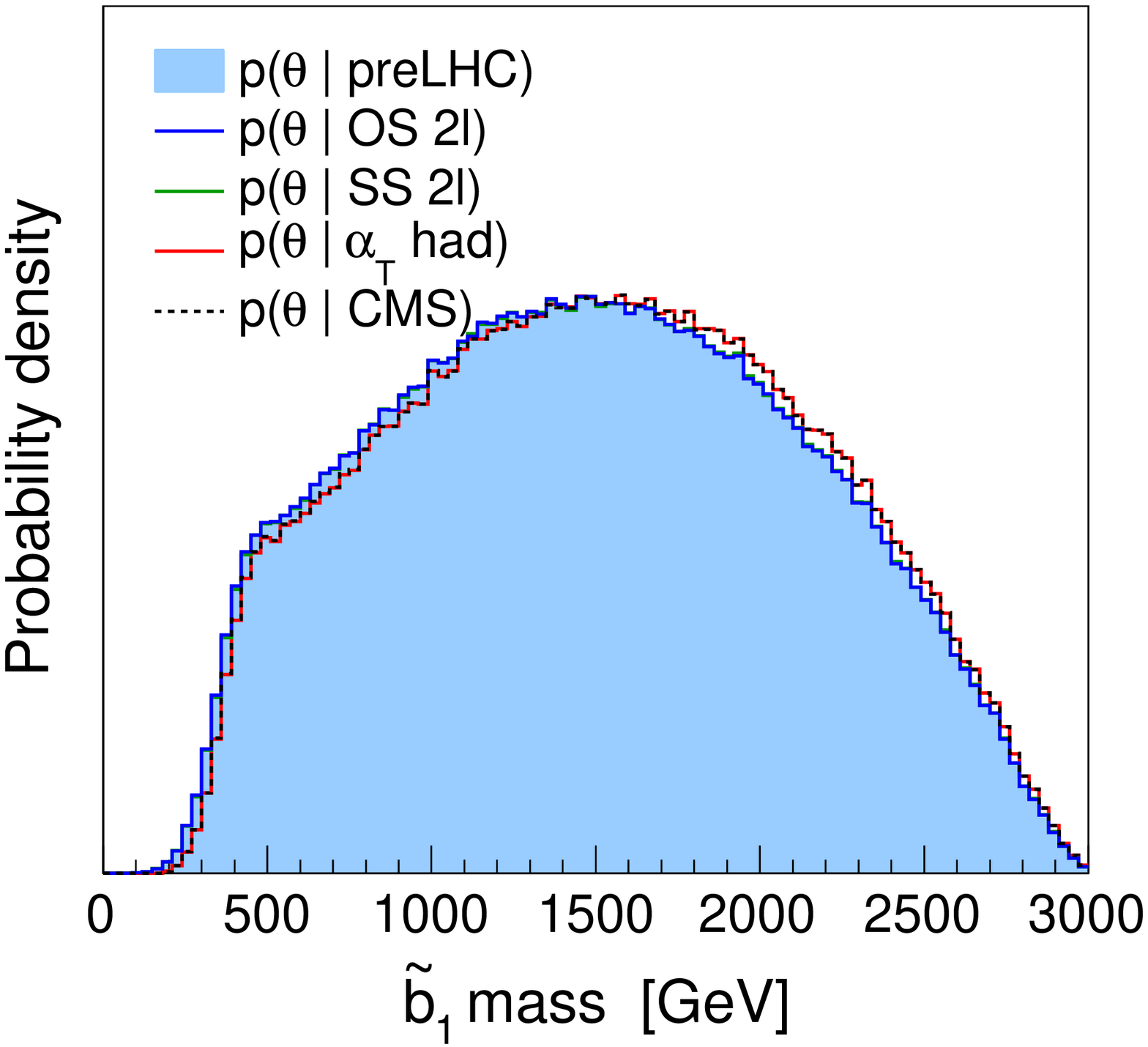} 
   \includegraphics[width=3.5cm]{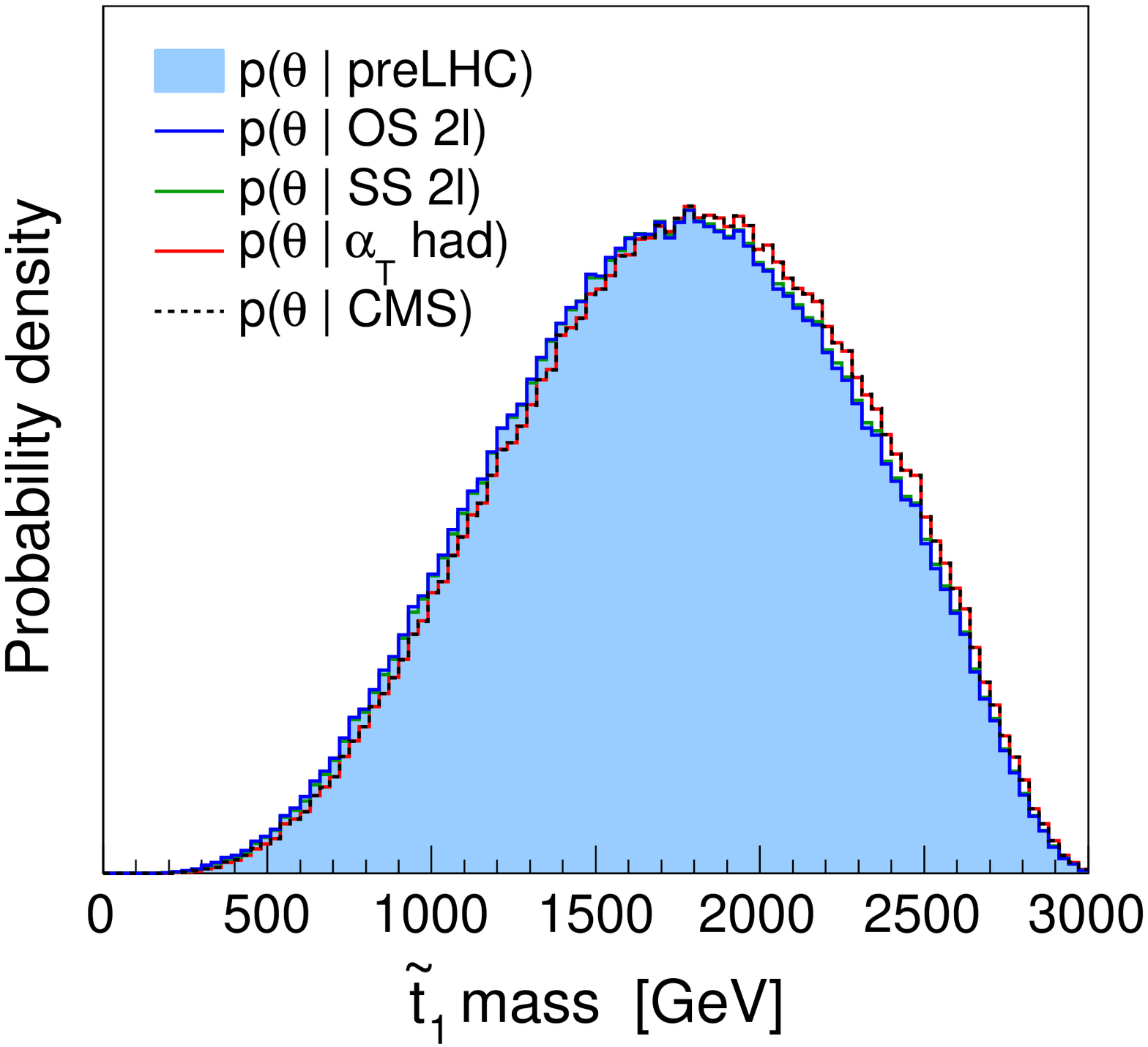} 
   \includegraphics[width=3.5cm]{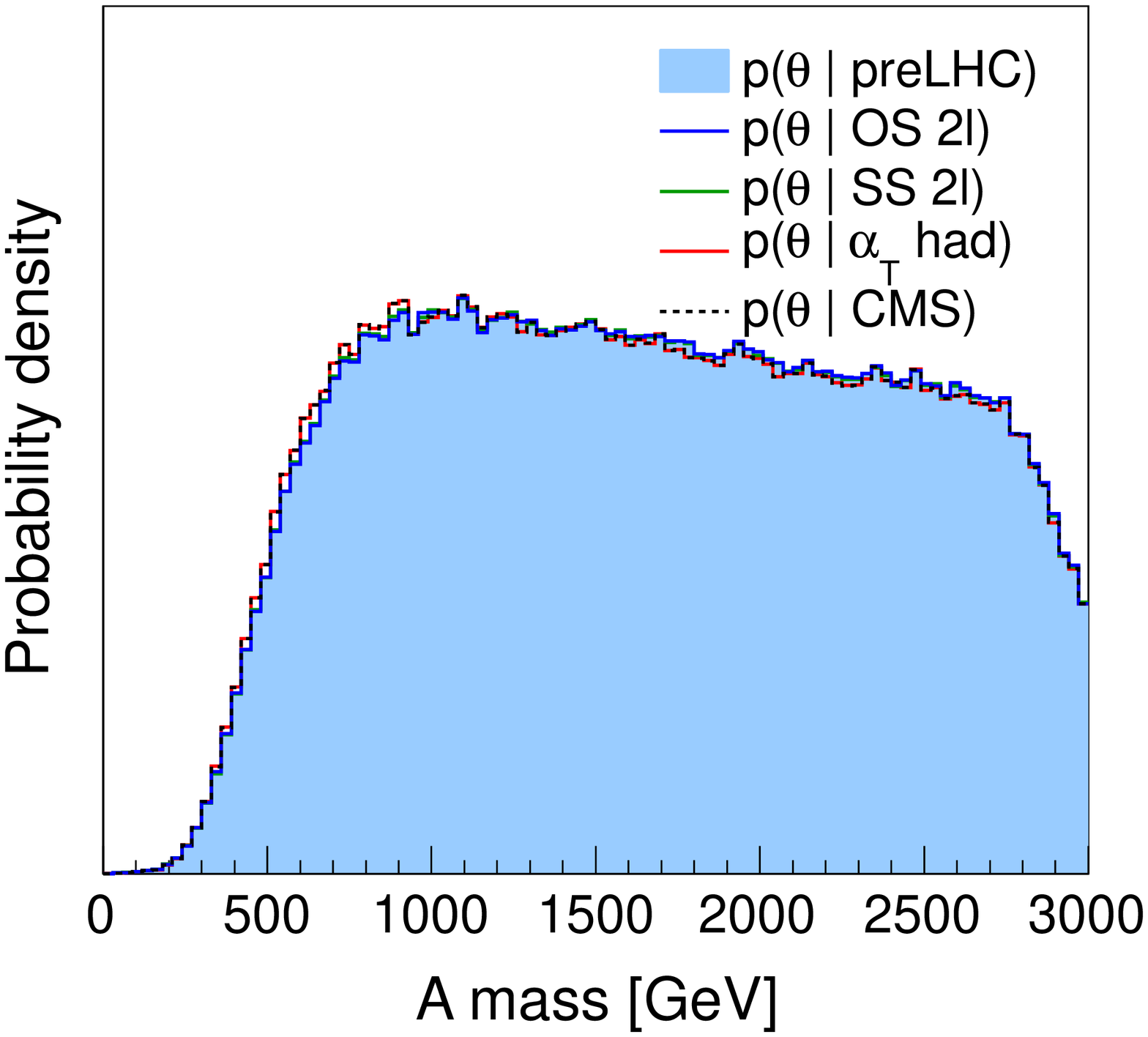} 
   \includegraphics[width=3.5cm]{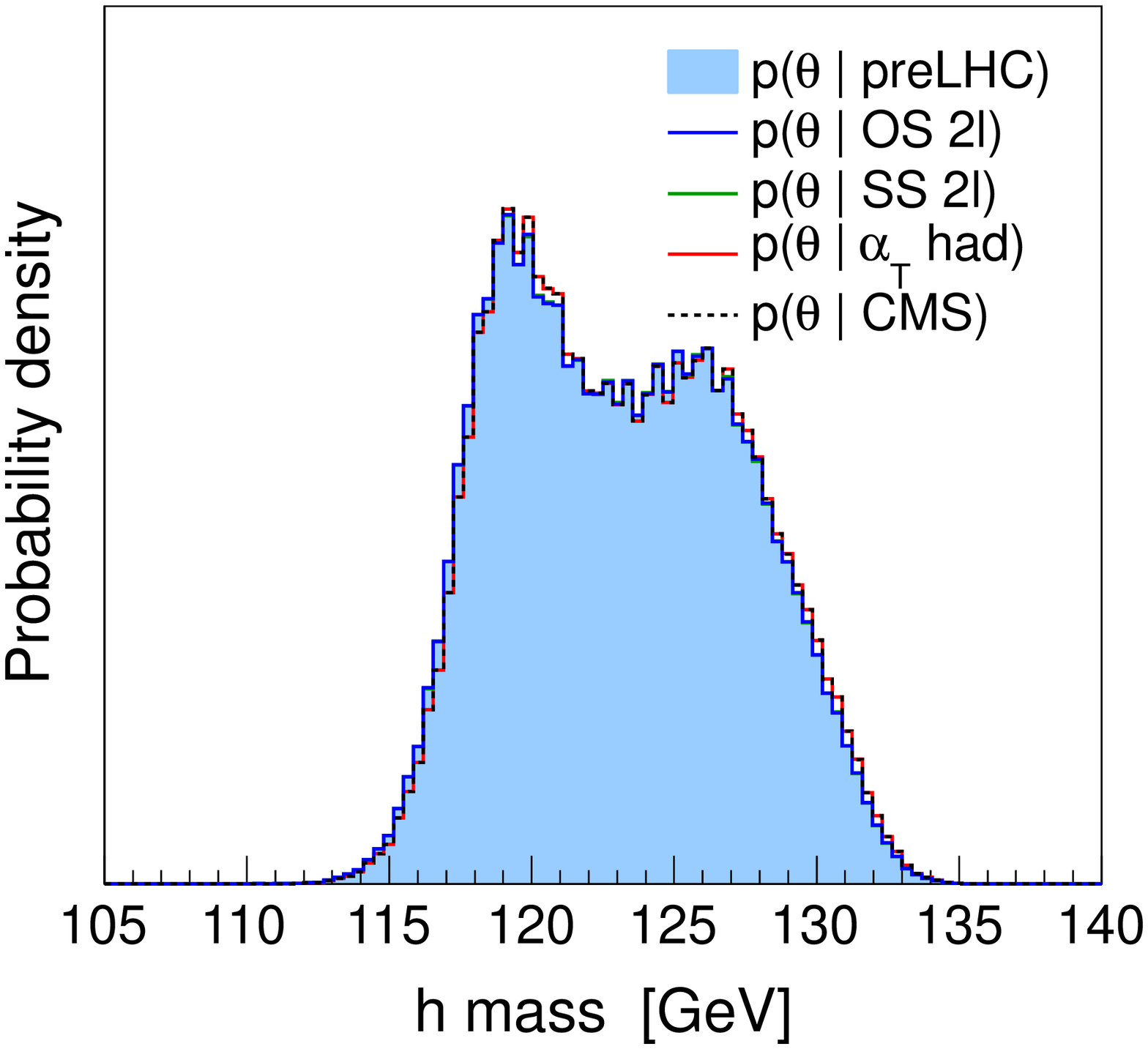} 
   \caption{Marginalized 1D posterior densities of sparticle and Higgs masses.}
   \label{fig:dist-1d-masses}
\end{figure}

We now present the results of this analysis.
Figure~\ref{fig:dist-1d-masses} shows marginalized 1-dimensional (1D) posterior probability density functions  of various sparticle and Higgs masses. The light blue histograms represent the preLHC probability densities, 
{\it i.e.}\ taking into account only the data listed in Table~\ref{tab:PLMs}. 
Note that the $\tilde\chi^\pm_1$ and $\tilde e_L/\tilde \mu_L$ are bound to be light by the $\Delta a_\mu$ constraint. Note also that our preLHC distributions differ somewhat from those presented in Ref.~\cite{AbdusSalam:2009qd} as we have chosen not to impose any constraint on $\Omega h^2$.

The blue, green and red lines show, respectively, the effects of the OS di-lepton, SS di-lepton and  
$\alpha_T$ hadronic CMS analyses. The dashed black lines show the final 
posterior densities after inclusion of the results of all three analyses. 
It is evident that with current LHC data-sets, the di-lepton analyses have very little effect on the 
posterior densities, while the $\alpha_T$ hadronic 
analysis pushes the gluino and $1^{st}/2^{nd}$-generation squark masses towards higher values. 
We also note the slight effect on the $\tilde\chi^0_1$ LSP mass. The masses of other sparticles, 
including charginos, sleptons and $3^{rd}$-generation squarks,  are basically unaffected by the 
current LHC results. This contrasts with the CMSSM case, in which all these masses are correlated 
through their dependence on $m_{1/2}$ and $m_0$. Finally, we see that the Higgs mass distributions, 
including that of $m_h$, remain unaffected by current SUSY searches. 

\begin{figure}[p]
   \centering
   \includegraphics[width=3.5cm]{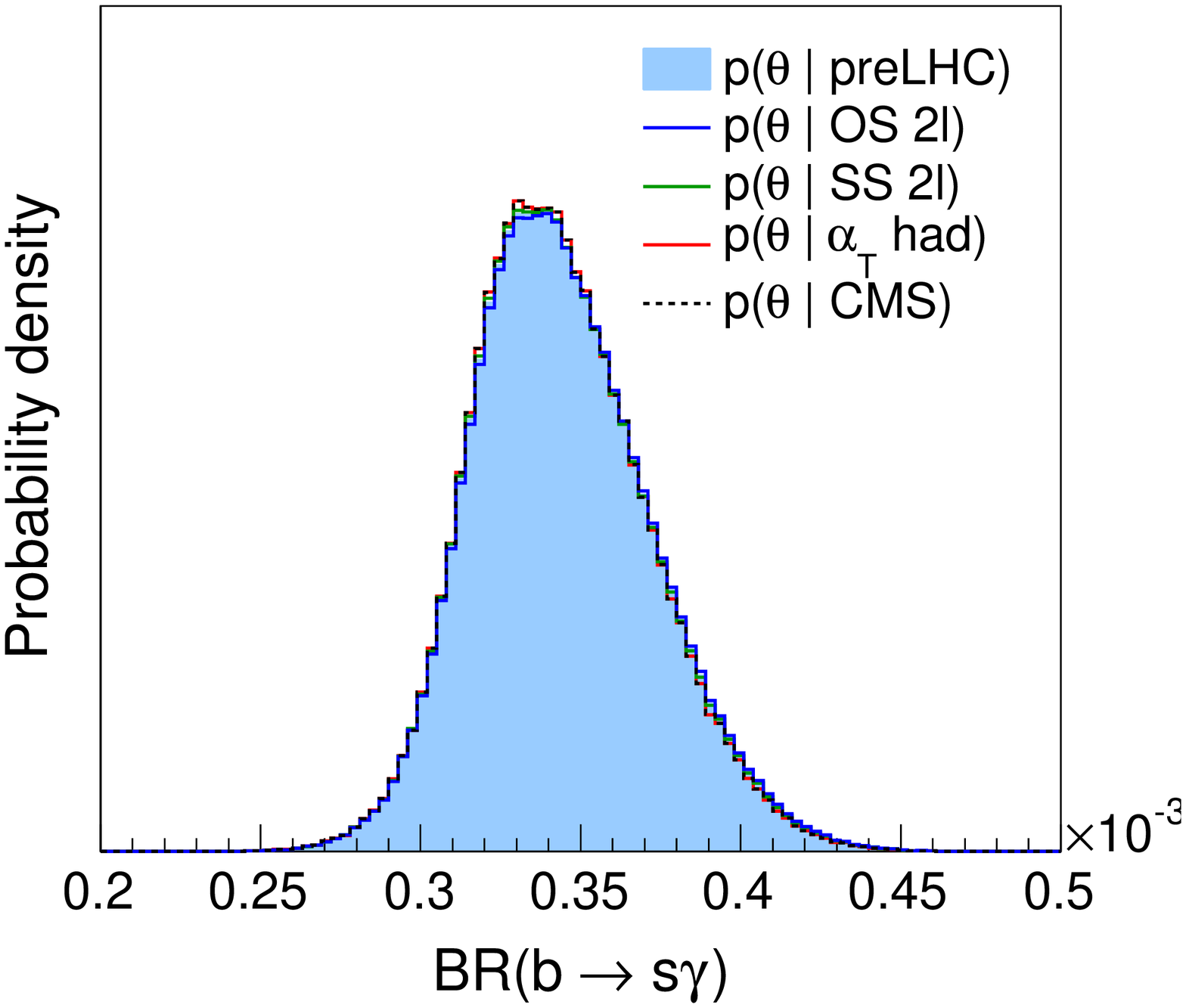} 
   \includegraphics[width=3.5cm]{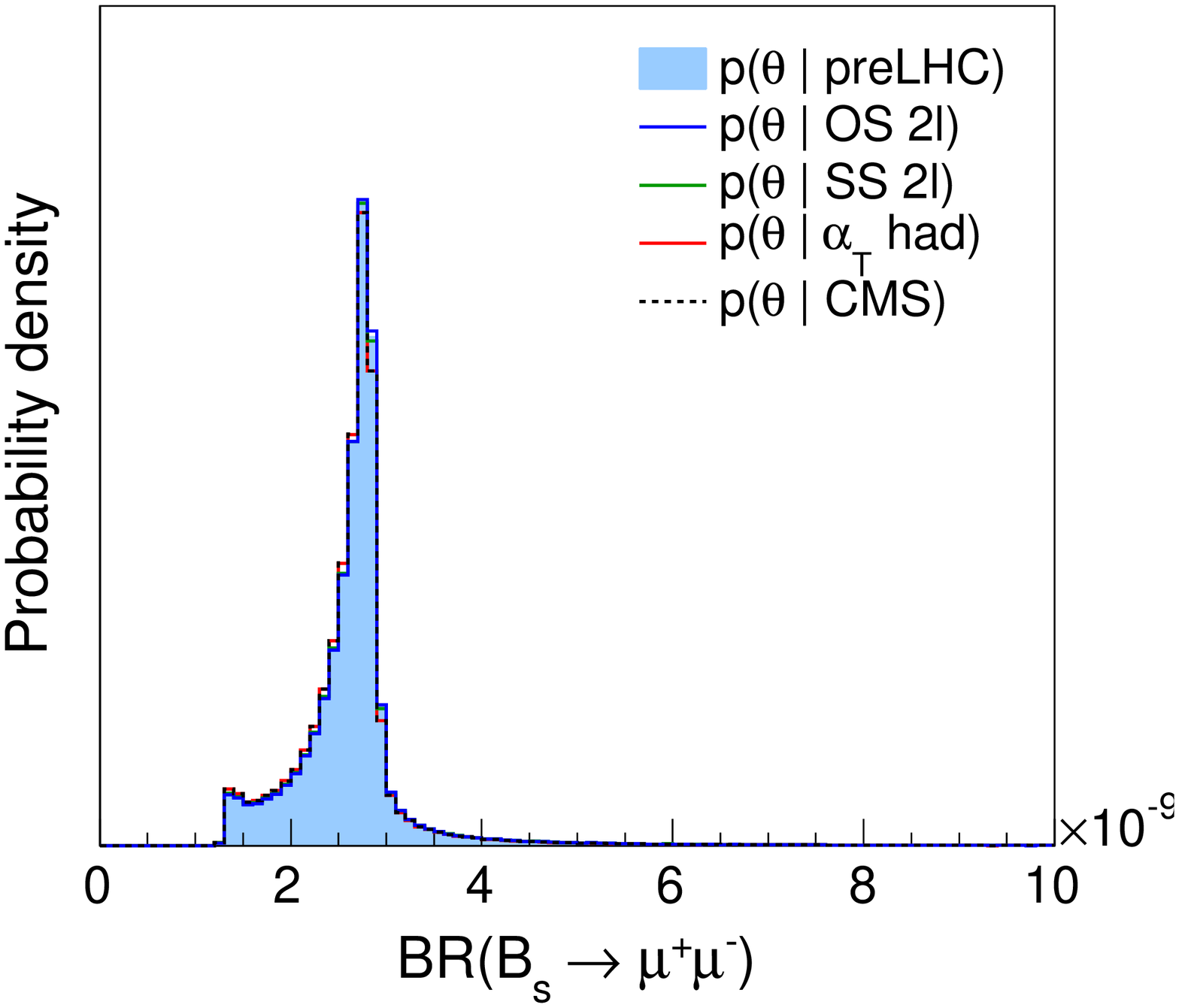} 
   \includegraphics[width=3.5cm]{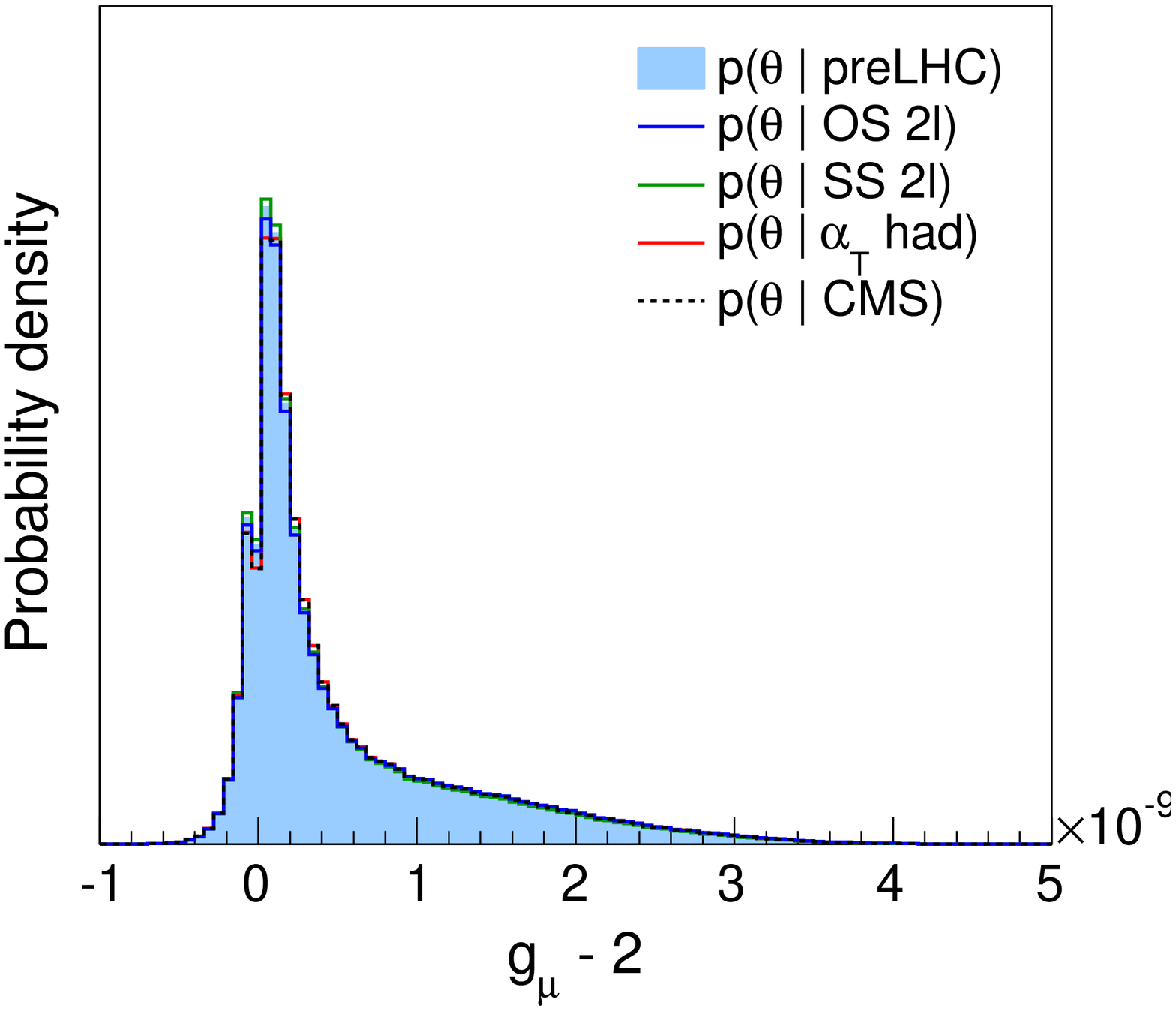} 
   \includegraphics[width=3.5cm]{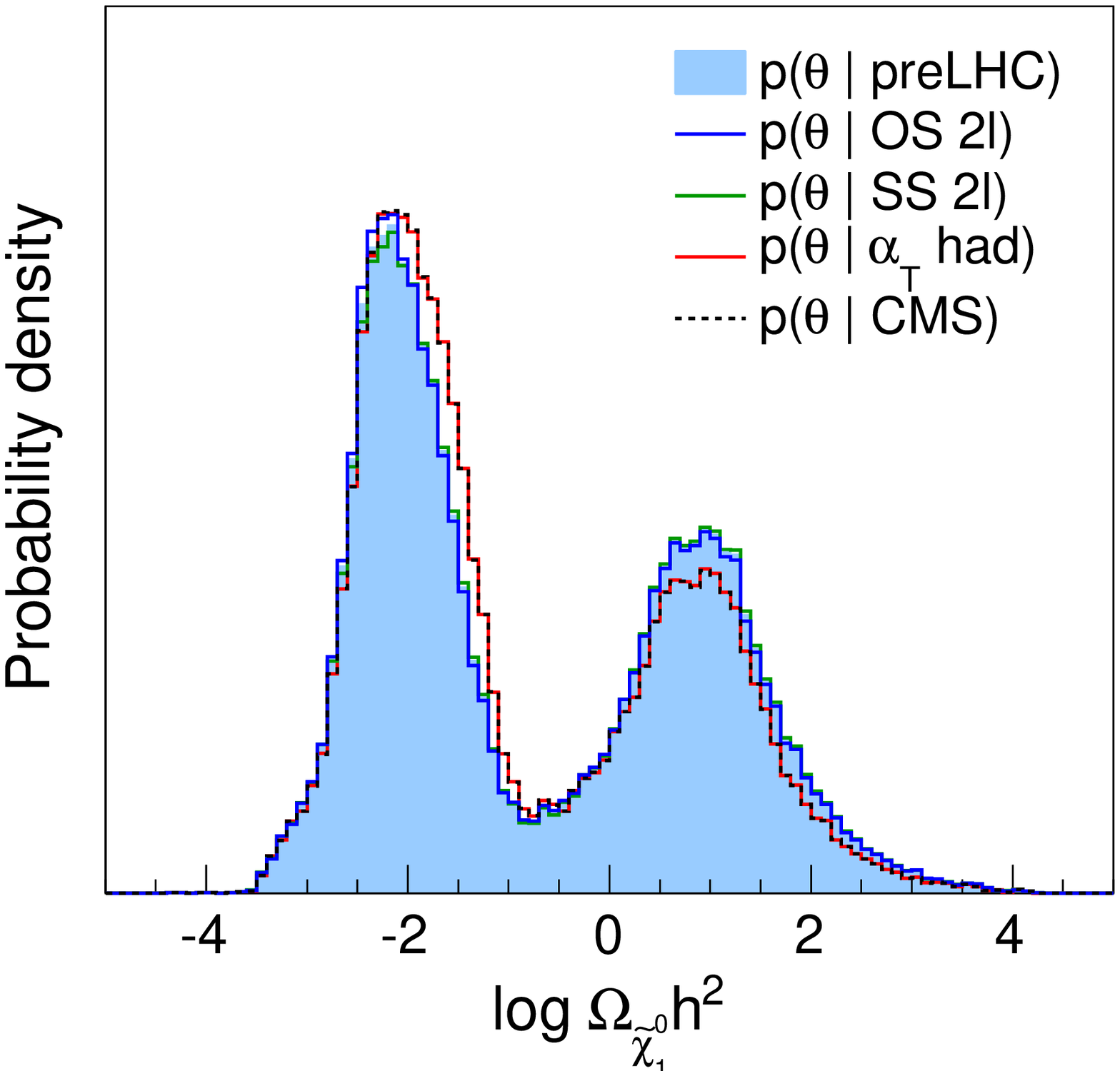} 
   \caption{Marginalized 1D posterior densities of $BR(b\to s\gamma)$,  $BR(B_s\to \mu^+\mu^-)$, SUSY contribution to $(g-2)_\mu$, and neutralino relic density $\Omega h^2$.}
   \label{fig:dist-1d-obs}
\end{figure}

\begin{figure}[p]
   \centering
   \includegraphics[width=3.5cm]{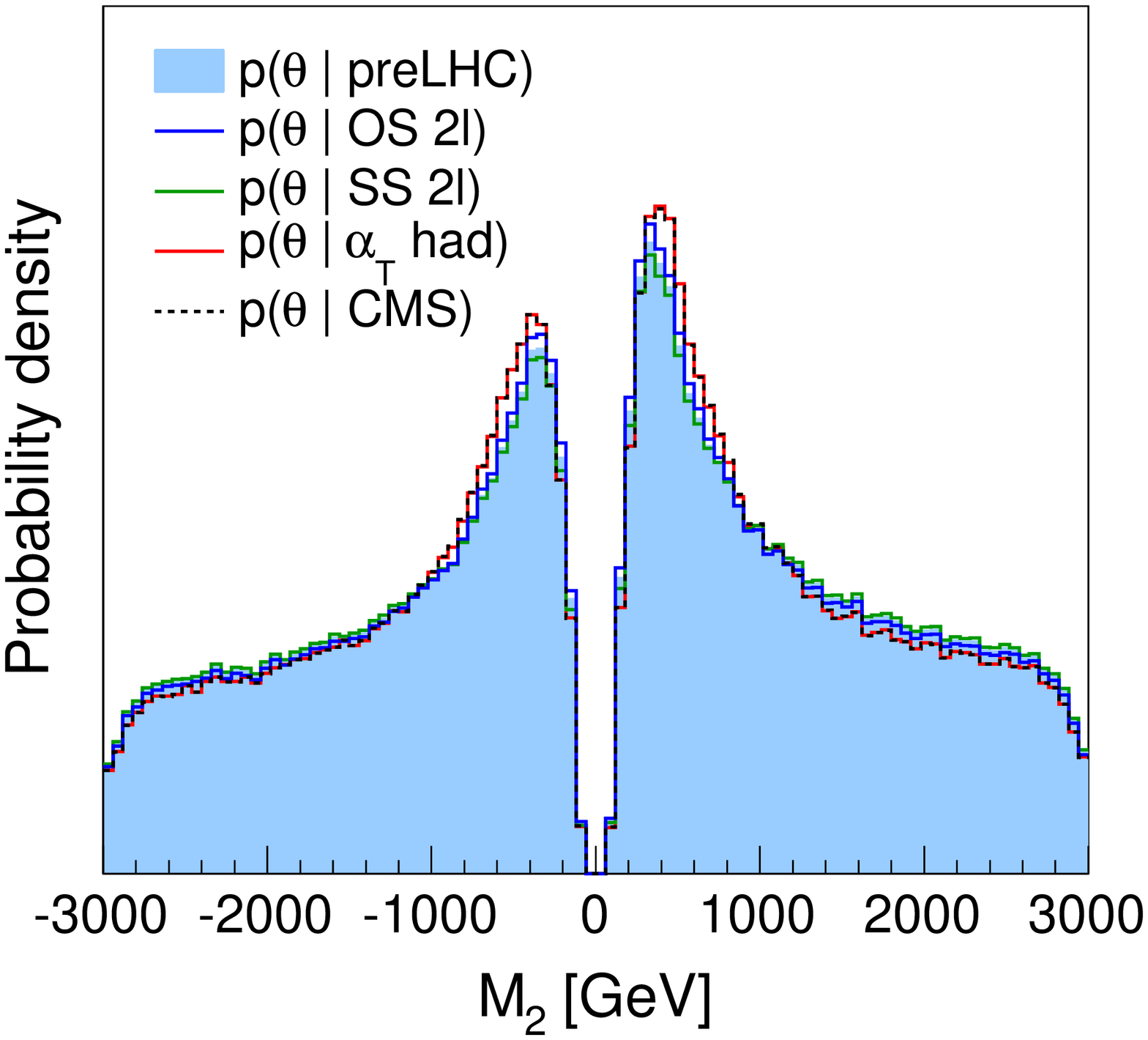} 
   \includegraphics[width=3.5cm]{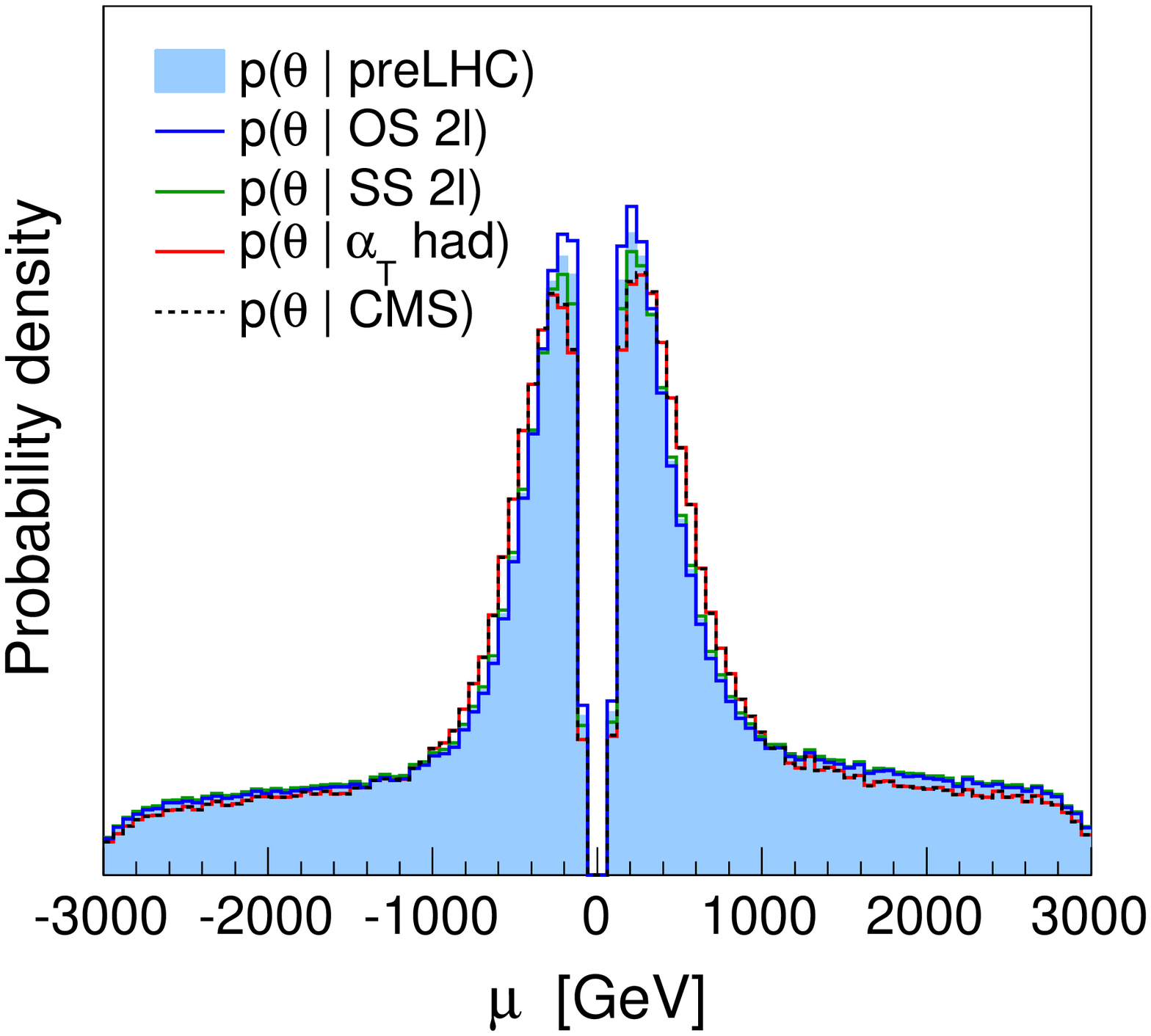} 
   \includegraphics[width=3.5cm]{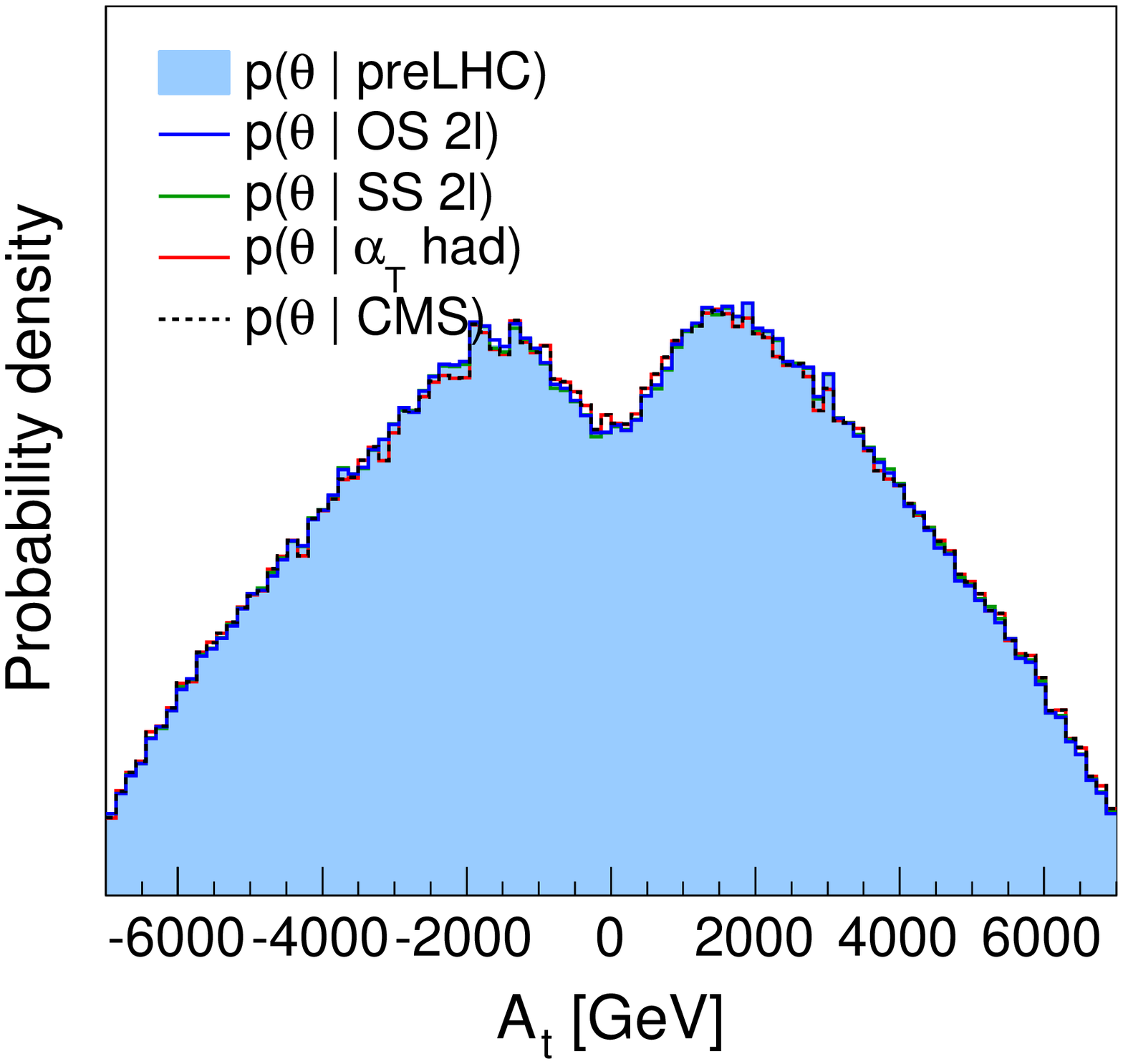} 
   \includegraphics[width=3.5cm]{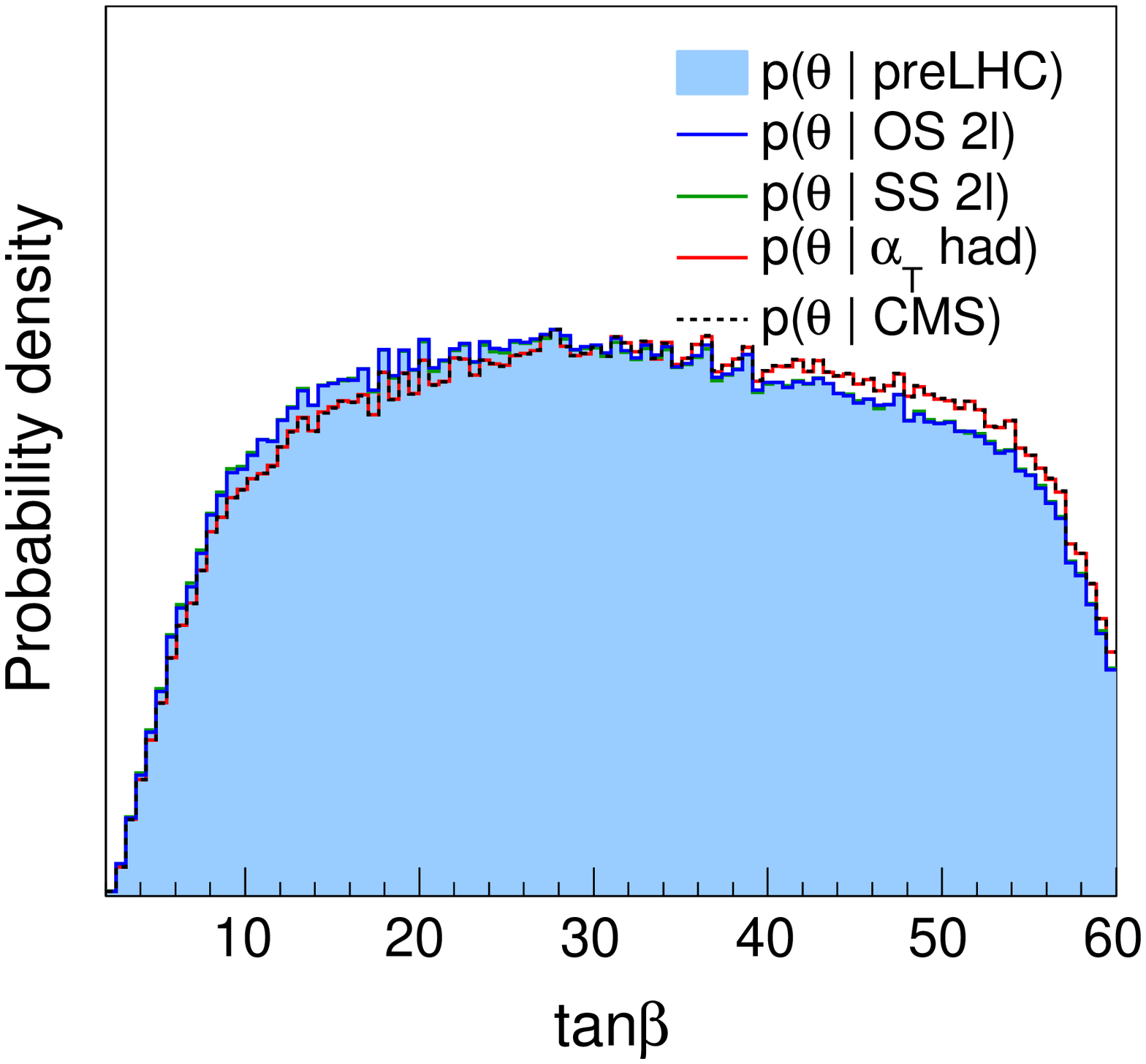} 
   \caption{Marginalized 1D posterior densities for $M_2$, $\mu$, $A_t$ and $\tan\beta$.}
   \label{fig:dist-1d-params}
\end{figure}

The 1D distributions of $BR(b\to s\gamma)$,  $BR(B_s\to \mu^+\mu^-)$, $\Delta a_\mu$ and the neutralino relic density $\Omega h^2$ are shown in Fig.~\ref{fig:dist-1d-obs}. 
In Fig.~\ref{fig:dist-1d-params} we show the posterior densities 
of some SUSY Lagrangian parameters: $M_2$, $\mu$, $A_t$ and $\tan\beta$. We observe 
a slight preference for $\mu>0$ with $p(\mu>0)\approx0.53$, both pre- and post-LHC startup. 
This is, however, inconclusive (as is the preference for $\mu<0$ found in \cite{AbdusSalam:2009qd}). 
We  confirm the sign correlations between $M_i$ and $\mu$, and between $A_t$ and $\mu$, already demonstrated in Ref.~\cite{AbdusSalam:2009qd}. 
The corresponding plots are available at Ref.~\cite{webpage}.


\begin{figure}[t]
   \centering
   \includegraphics[width=12cm]{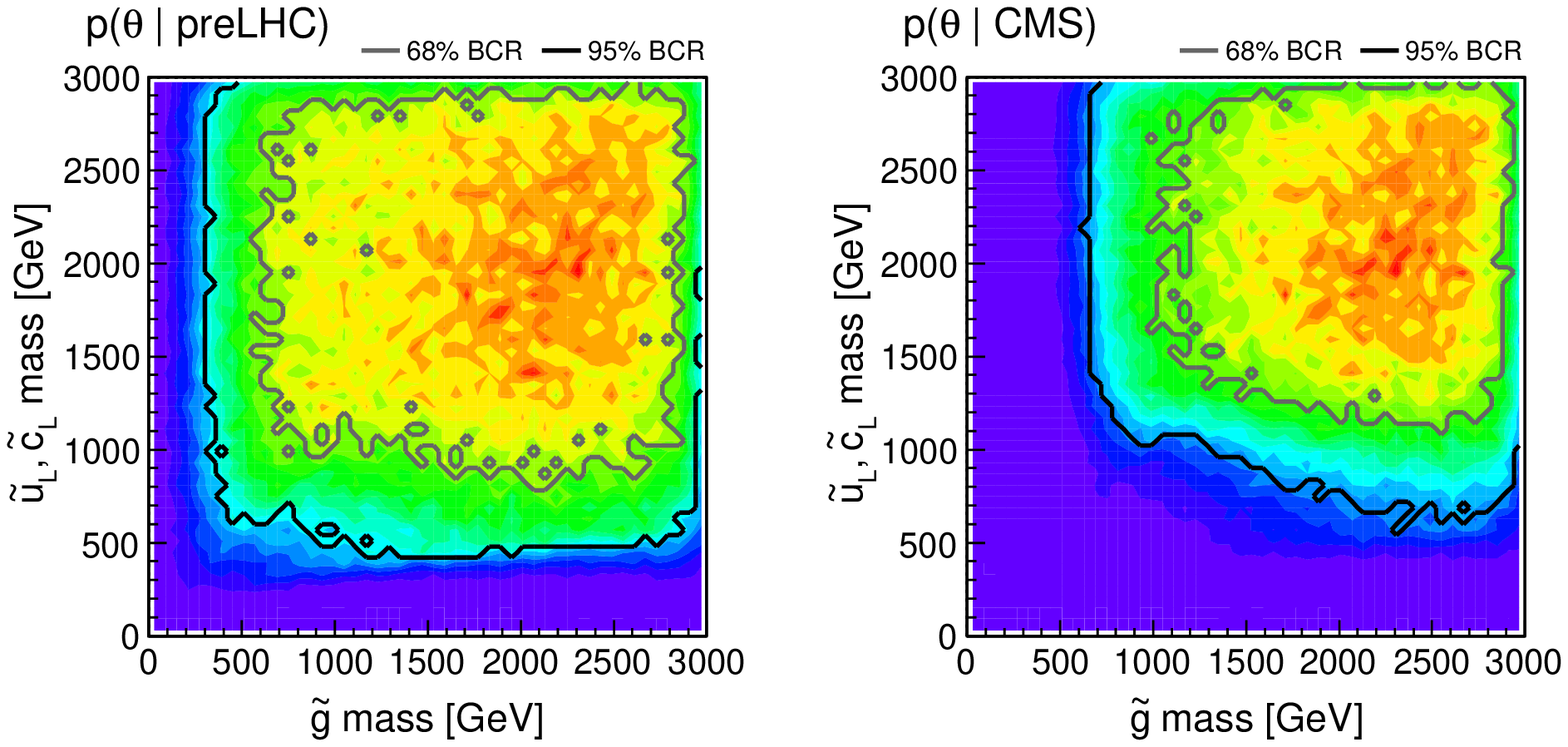}\\
   \includegraphics[width=12cm]{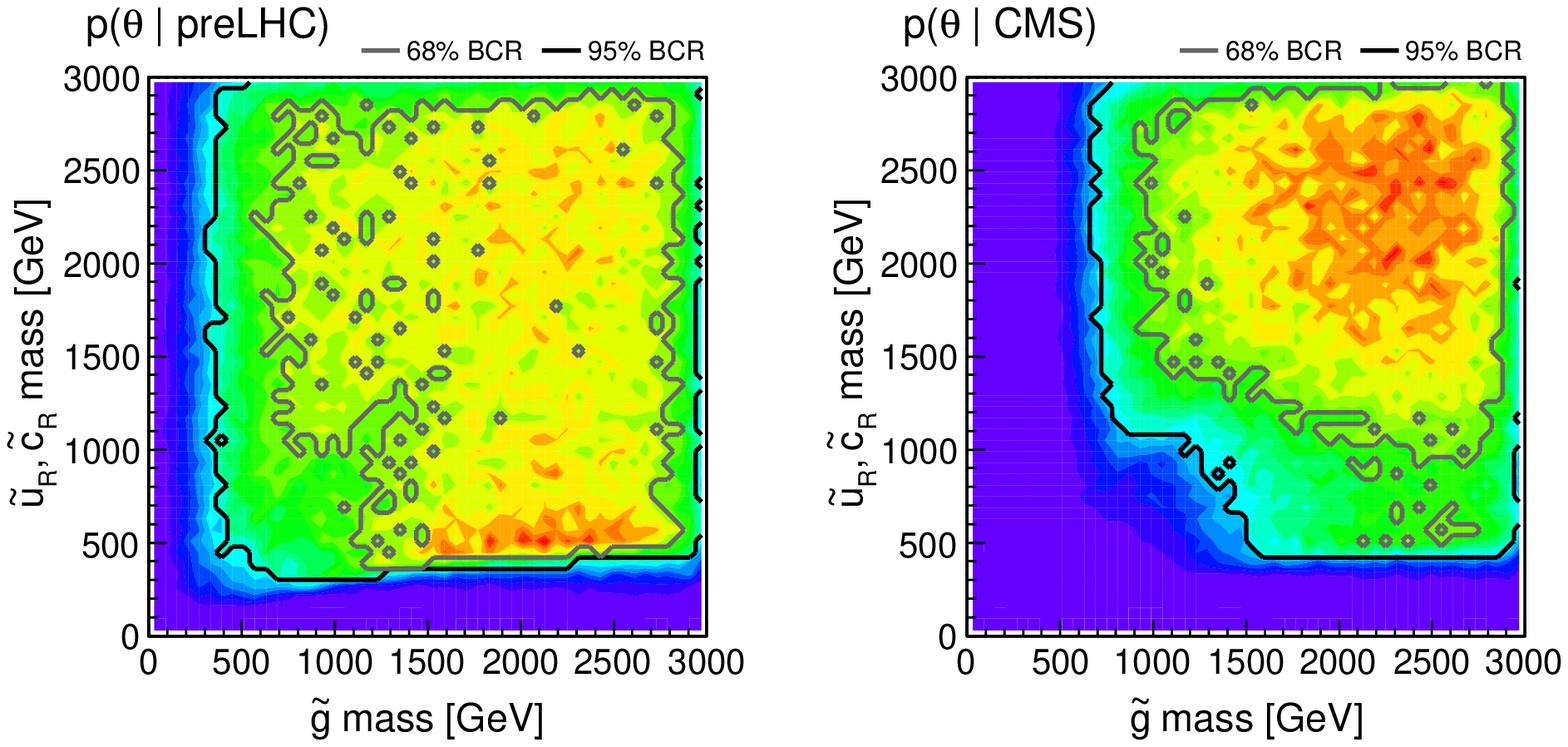} 
   \caption{Marginalized 2D posterior densities of gluino versus squark masses, on the left before and on the right after taking the CMS searches into account. The grey and black contours enclose the 68\% and 95\% Bayesian credible regions, respectively.}
   \label{fig:dist-2d-mg-msquark}
\end{figure}

It is also instructive to consider correlations between different sparticle masses. 
Figure~\ref{fig:dist-2d-mg-msquark} demonstrates the impact of the CMS analyses in the 
$(m_{\tilde u_L}, m_{\tilde g})$ and $(m_{\tilde u_R}, m_{\tilde g})$ planes. 
It is interesting to note that the boundaries of the LHC 95\% Bayesian credible regions (BCRs) approximately match the 95\%\,CL exclusion limits in the CMSSM.\footnote{The way to calculate the BCRs is to some 
extent a matter of choice.Ê Here we select the region containing the highest posterior density values.Ê 
This is equivalent to choosing the minimal area that contains the 68\% or 95\% of the total volume.} 
We deduce that $m_{\tilde g,\tilde q}\gtrsim 1.1$~TeV for  
$m_{\tilde g}\simeq m_{\tilde q}$, and $m_{\tilde g}\gtrsim 700$~GeV for $m_{\tilde q}\gg m_{\tilde g}$, is a  robust conclusion that persists beyond the CMSSM or Simplified Models.\footnote{The characteristics of 
the pMSSM points with $m_{\tilde g}<700$~GeV that are \emph{not} excluded by the current 
SUSY analyses will be the subject of a subsequent study; see also 
Refs.~\cite{Conley:2011nn,AbdusSalam:2011hd} in this context.}

\begin{figure}[t]
   \centering
   \includegraphics[width=12cm]{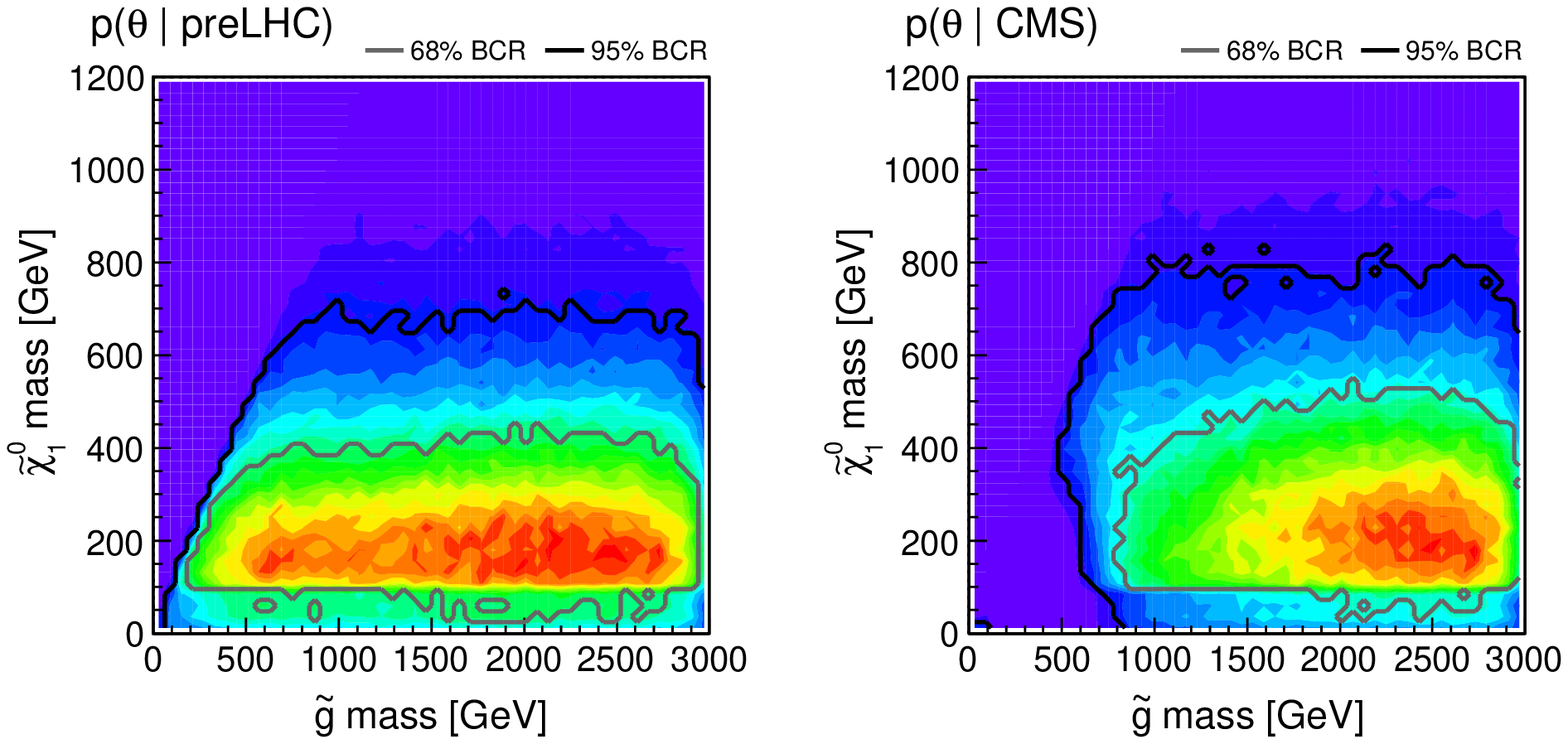}\\ 
   \includegraphics[width=12cm]{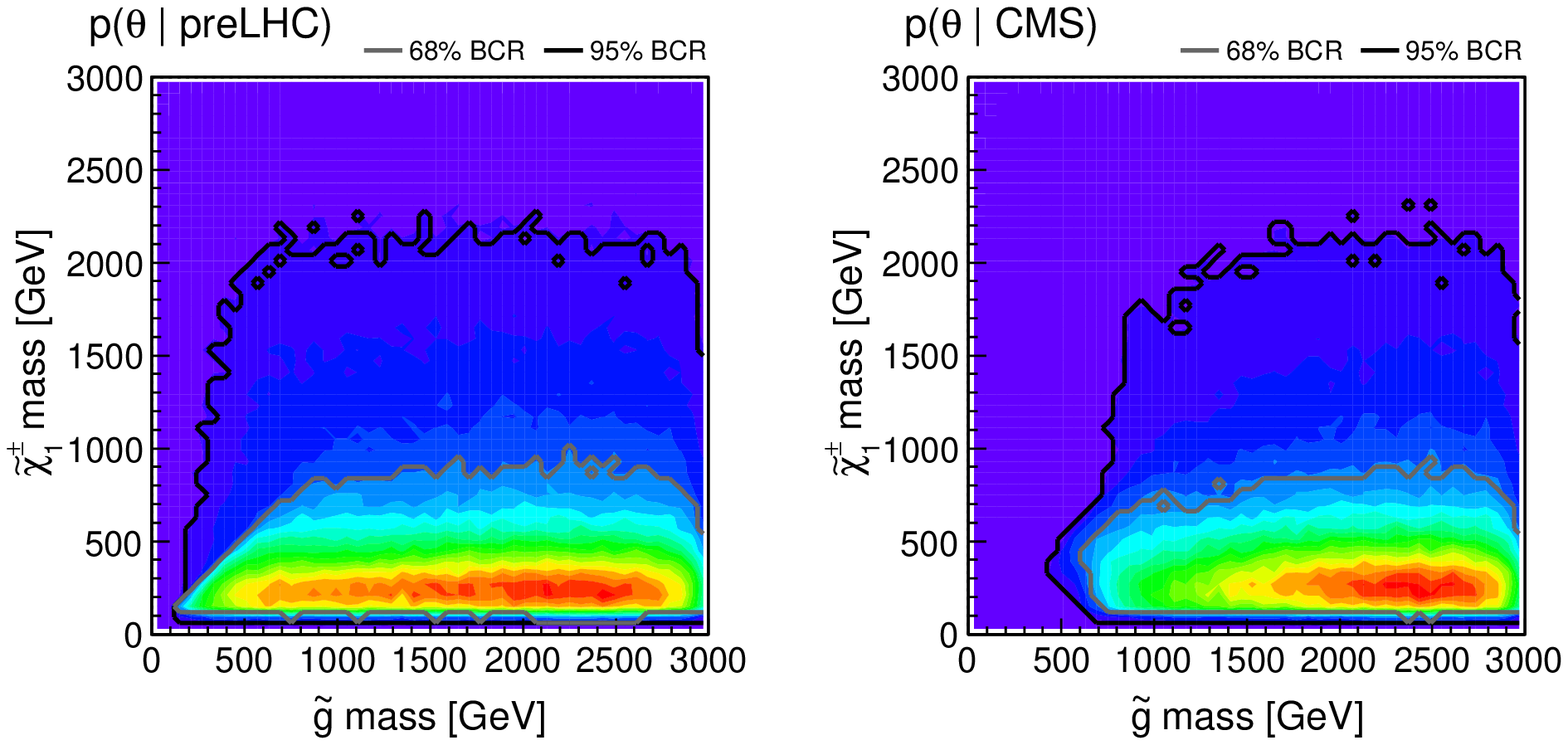} 
   \caption{Marginalized 2D posterior densities of gluino versus neutralino and of gluino versus chargino mass, on the left before and on the right after taking the CMS searches into account. The grey and black contours enclose the 68\% and 95\% Bayesian credible regions, respectively.}
   \label{fig:dist-2d-mg-mw1}
\end{figure}

Our approach moreover allows the study of dependencies between other masses  in a straightforward way, 
as illustrated in Fig.~\ref{fig:dist-2d-mg-mw1} by means of posterior densities in the 
$(m_{\tilde\chi^0_1}, m_{\tilde g})$ and $(m_{\tilde\chi^\pm_1}, m_{\tilde g})$ planes. 
We now see explicitly that bounds on the gluino mass are not reflected in chargino and neutralino masses, as would be the case in the CMSSM (or actually any scheme with gaugino-mass universality). Moreover, such plots permit other interesting observations. 
In particular, we see how the sensitivity of CMS searches to the gluino mass 
worsens for increasing neutralino or chargino mass. 
Additional plots of 1D and 2D distributions are available at Ref.~\cite{webpage}.

At this point a comment is in order regarding prior dependence.  
Our preLHC distributions are of course subject to the same prior dependence 
that was discussed in Ref.~\cite{AbdusSalam:2009qd}, and persists for quantities that 
are not much affected by the CMS measurements. 
However, we expect that as the effect of data becomes more influential, the sensitivity 
to the prior diminishes. In fact, for the gluino and $1^{st}/2^{nd}$ generation squark masses, 
the likelihood based on LHC data already dominates the prior. For these quantities the 
posterior distributions are indeed found to be insensitive to the choice of priors. (This is also true  for 
some other  quantities such as the $\tilde t_1$ and $h^0$ masses,
which are already well-constrained by preLHC data.) 

\begin{figure}[t]
   \centering
   \includegraphics[width=6cm]{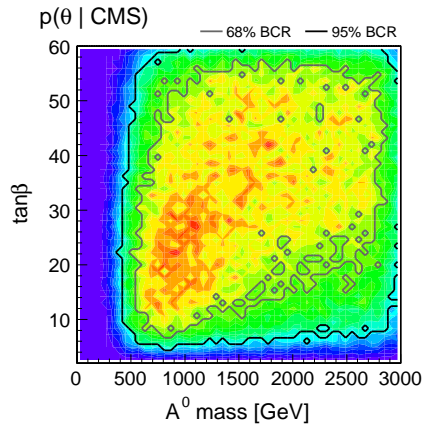} 
   \caption{Marginalized 2D posterior density of $\tan\beta$ versus $m_A$. The grey and black contours enclose the 68\% and 95\% Bayesian credible regions, respectively. 
   The preLHC distribution looks essentially the same.}
   \label{fig:dist-2d-mAtb}
\end{figure}

It is also interesting to consider the interplay with other, non-SUSY, searches. 
Regarding the Higgs sector, in particular the results on $H/A\to\tau\tau$ may have some impact 
on the pMSSM global fit. ATLAS and CMS searches for $H/A\to\tau\tau$ currently exclude 
$\tan\beta\lesssim10$--20 for $m_A\lesssim 250$~GeV, and $\tan\beta\lesssim50$--60 for 
$m_A=450$--500~GeV~\cite{nikitenko}.
While it will be interesting to include this in our global analysis, we note that our 95\% BCR in the ($\tan\beta, m_A$) plane, displayed in Fig.~\ref{fig:dist-2d-mAtb}, starts at $m_A\approx 500$~GeV and shows no significant dependence on $\tan\beta$. 

\begin{figure}[t]
   \centering
   \includegraphics[width=12cm]{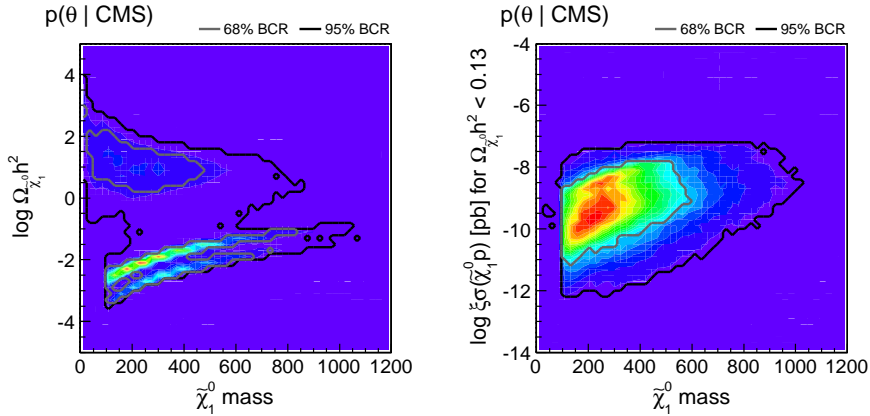} 
   \caption{Marginalized 2D posterior densities of $\Omega h^2$ (left) 
   and rescaled spin-independent scattering cross section off protons (right) 
   versus LSP mass. For the latter, only points with $\Omega h^2<0.13$ are taken into account.    The grey and black contours enclose the 68\% and 95\% Bayesian credible regions, respectively.}
   \label{fig:dist-2d-dark}
\end{figure}

Finally, we illustrate in Fig.~\ref{fig:dist-2d-dark} the interplay with dark matter searches. 
On the left, we show the posterior density in the ($\Omega_\chi h^2,m_{\tilde\chi^0_1}$) 
plane. While matching the cosmologically observed value 
$\Omega h^2=0.1123\pm0.0035$~\cite{Jarosik:2010iu}
needs a high degree of fine tuning, a $\tilde\chi^0_1$ that is at least part of the dark matter 
has a probability of about 60\%; see also the right-most plot in Fig.~\ref{fig:dist-1d-obs}.
In fact there is a slight increase from $p(\Omega_\chi h^2<0.123)=0.53$ with preLHC data 
to $p(\Omega_\chi h^2<0.123)=0.59$ when including the CMS analyses, scarcely depending 
on the exact value of the upper bound on $\Omega_\chi h^2$. 
The right plot in Fig.~\ref{fig:dist-2d-dark} shows the posterior density 
of the spin-independent scattering cross section off protons, $\xi\sigma^{\rm SI}(\tilde\chi^0_1p)$, 
for the case that the LSP is at least part of the dark matter. Here we imposed $\Omega h^2<0.13$ 
and rescaled the cross section by a factor $\xi=\Omega h^2/0.1123$. 
Note that the most credible region is yet to be tested by the direct dark matter searches. 

\section{Conclusions}\label{sec:conclusions}

We presented the first interpretation of the 2011 LHC results based on $\sim$1\,fb$^{-1}$ of data within the framework of the phenomenological MSSM---a sufficiently generic and well-motivated 19-dimensional parameterization of SUSY defined at the SUSY scale.  We have used three independent LHC SUSY analyses, namely, the CMS $\alpha_T$ hadronic, opposite-sign dilepton and same-sign di-lepton analyses for this purpose, and expressed our results in terms of posterior probability densities.

Our bounds on gluino and 1$^{st}$/2$^{nd}$-generation squark masses match those derived in the CMSSM by the experimental collaborations. In addition, we were able to make independent statements on the masses and properties of the other SUSY (and Higgs) particles, and to show relations between masses that weaken the current bounds. 
In the 
chargino-versus-gluino-mass plane, for instance, the boundary of the 95\% BCR 
can go down from $m_{\tilde g}\approx 800$~GeV to $m_{\tilde g}\approx 400$~GeV, depending on 
the $\tilde\chi^\pm_1$ mass and the rest of the spectrum. 
Our results thus show that current SUSY searches at the LHC provide rather limited constraints on supersymmetry in general. With the currently available data and searches, we have indeed been able to probe only a small portion of the vast pMSSM parameter space, while many regions are still waiting to be explored.  Being able to work constructively with generic multi-parameter models such as pMSSM will serve as a guide to identify the unexplored regions and devise a broader range of dedicated searches sensitive to these.

We have demonstrated in this study that the interpretation of LHC results in terms of broad classes of multi-parameter SUSY models is feasible with the currently available computational and statistical tools, and that it is indeed possible to make meaningful statements on the nature of such models and therefore on supersymmetry, in general.
It will be interesting to extend our study to include also results from non-SUSY searches. 
Indeed, one of the major advantages of our approach is that it is very well suited for global analyses of  multiple results from the LHC and elsewhere.

\section*{Acknowledgements}

We thank J.~Hewett and T.~Rizzo for discussions on ``SUSY without Prejudice'' and related technical issues. 
Moreover, we thank F.~Mahmoudi and K.~Williams for help with interfacing {\tt HiggsBounds}, and 
M.~M\"uhlleitner for fixing {\tt SUSYHIT}.
This work was supported in part by the U.S. Department of Energy under Grant No.
DE-FG02-97ER41022 and by IN2P3 under grant PICS FR-USA 5872.


\end{document}